\documentclass{article}

\usepackage{amsfonts,url}
\usepackage{amssymb}
\usepackage{amsmath}
\usepackage{amstext,verbatim,graphicx,ifthen,multirow,amsthm}
\usepackage{bm}

\usepackage{graphicx}
\usepackage{verbatim}
\usepackage{caption}
\usepackage{subcaption}
\usepackage{fancyvrb}
\usepackage{enumerate}
\usepackage{setspace}

\usepackage{refcount}

\usepackage{relsize}
\usepackage{float}

\usepackage{natbib}

\theoremstyle{plain}

\newtheorem*
{theo*}{Theorem}

\newtheorem{prop}{Proposition}

\newtheorem{coro}[prop]{Corollary}
\newtheorem{lemm}[prop]{Lemma}
\newtheorem{theo}[prop]{Theorem}

\theoremstyle{definition}

\newtheorem{assu}{Assumption}

\usepackage{guido_tex}
\usepackage{stefan_tex}

\newcommand{\mutt}{\mu_{\rm t}}
\newcommand{\hmutt}{\hat{\mu}_{\rm t}}
\newcommand{\hmucc}{\hat{\mu}_{\rm c}}
\newcommand{\mucc}{\mu_{\rm c}}
\newcommand{\betatt}{\beta_{\rm t}}
\newcommand{\hbetatt}{\hat{\beta}_{\rm t}}
\newcommand{\betacc}{\beta_{\rm c}}
\newcommand{\hbetacc}{\hat{\beta}_{\rm c}}

\newcommand{\bxcc}{{\bf X}_{\rm c}}
\newcommand{\bxtt}{{\bf X}_{\rm t}}
\newcommand{\haty}{\overline{Y}_{\rm t}}
\newcommand{\hatx}{\overline{X}_{\rm t}}
\newcommand{\hatxj}{\overline{X}_{{\rm t},j}}
\newcommand{\bX}{\overline{X}}

\theoremstyle{definition}
\newfloat{algbox}{tbp}{lop}
\newtheorem{alg}{Procedure}
\newcommand{\myalg}[3]{
\begin{center}
\fbox{
\parbox{0.95\textwidth}{
\begin{alg}\label{#1}{\textsc{ #2}}
\vspace{.1cm}\\ #3 
\end{alg}
}}
\end{center}
}

\usepackage{hyperref}
\usepackage[margin=1.1in]{geometry}
\hypersetup{colorlinks,citecolor=blue,urlcolor=blue,linkcolor=blue}

\newcommand{\he}{\hat{e}}

\newcommand{\bycc}{{\bf Y}_\tc}

\newcommand{\hmuC}{\hmucc}

\begin{document}

\title{Approximate Residual Balancing: De-Biased Inference of Average Treatment Effects in High Dimensions\thanks{
We are grateful for detailed comments from
Jelena Bradic, Edgar Dobriban, Bryan Graham, Chris Hansen, Nishanth Mundru, Jamie Robins and Jos{\'e} Zubizarreta,
and for discussions with seminar participants at
the Atlantic Causal Inference Conference,
Boston University,
the Columbia Causal Inference Conference,
Columbia University,
Cowles Foundation, 
the Econometric Society Winter Meeting,
the EGAP Standards Meeting,
the European Meeting of Statisticians,
ICML,
INFORMS,
Stanford University,
UNC Chapel Hill,
University of Southern California,
and the World Statistics Congress.}}

\author{
Susan Athey\thanks{{Professor of Economics, Stanford Graduate School of Business, and NBER, \texttt{athey@stanford.edu}. }}
\and Guido W. Imbens\thanks{{Professor of Economics, Stanford Graduate School of Business, and NBER,
\texttt{imbens@stanford.edu}. }}
 \and Stefan Wager\thanks{{Assistant Professor of Operations, Information and Technology and of Statistics (by courtesy), Stanford  Graduate School of Business, \texttt{swager@stanford.edu}. }} 
 }
\date{ Current version \ifcase\month\or
January\or February\or March\or April\or May\or June\or
July\or August\or September\or October\or November\or December\fi \ \number%
\year\ \  }
\maketitle

\begin{abstract}
There are many settings where researchers are interested in estimating average treatment effects and are willing to rely on the unconfoundedness assumption, which requires that the treatment assignment be as good as random conditional on pre-treatment variables.  The unconfoundedness assumption is often more plausible if a large number of pre-treatment variables are included in the analysis, but this can worsen the performance of standard approaches to treatment effect estimation. In this paper, we develop a method for de-biasing penalized regression adjustments to allow sparse regression methods like the lasso to be used for $\sqrt{n}$-consistent inference of average treatment effects in high-dimensional linear models. Given linearity, we do not need to assume that the treatment propensities are estimable, or that the average treatment effect is a sparse contrast of the outcome model parameters. Rather, in addition standard assumptions used to make lasso regression on the outcome model consistent under 1-norm error, we only require overlap, i.e., that the propensity score be uniformly bounded away from 0 and 1. Procedurally, our method combines balancing weights with a regularized regression adjustment.

\vspace{\baselineskip}

\end{abstract}

\section{Introduction}
\label{section:introduction}

In order to identify causal effects in observational studies, practitioners may assume
treatment assignments to be as good as random (or unconfounded) conditional on observed features of the units;
see \citet{rosenbaum1983central} and \citet{imbens2015causal} for general discussions.
Motivated by this setup, there is a large literature on how to adjust for differences in observed features
between the treatment and control groups; some popular methods include regression, matching,
propensity score weighting and subclassification, as well as doubly-robust combinations thereof
\citep[e.g.,][]{abadie, heckman1998matching, hirano2003efficient,
robins1994estimation, robins2, robins2008higher, rosenbaum2002observational, tan2010bounded, tsiatis2007semiparametric, van2006targeted}.

In practice, researchers sometimes need to account for a substantial number of features to make this assumption of unconfoundedness plausible. For example, in an observational study of the effect of flu vaccines on hospitalization, we may be concerned that only controlling for differences in the  age and sex distribution between controls and treated may not be sufficient to eliminate biases. In contrast, controlling for detailed medical histories and personal characteristics may make unconfoundedness more plausible.
But the formal asymptotic theory in the earlier literature only considers the case where the sample size increases while the number of features remains fixed, and so approximations based on those results may not yield valid inferences in settings where the number of features is large, possibly even larger than the sample size.

There has been considerable recent interest in adapting methods from the earlier literature to high-dimensional settings.
\citet{belloni2014inference,belloni2013program} show that
attempting to control for high-dimensional confounders using a regularized regression adjustment
obtained via, e.g., the lasso, can result in substantial biases.
\citet{belloni2014inference} propose an augmented variable selection scheme to avoid this effect,
while \citet{belloni2013program}, \citet{chernozhukov2016double},  \citet{farrell2015robust}, and \citet{van2011targeted}
build on the work of \citet{robins1994estimation,robins2} and discuss how a doubly robust approach
to average treatment effect estimation in high dimensions can also be used to compensate for the bias of regularized regression adjustments.
Despite the breadth of research on the topic, all the above papers rely crucially on the existence of a
consistent estimator of the propensity score, i.e., the conditional probability of receiving treatment given the features,
in order to yield $\sqrt{n}$-consistent estimates of the average treatment effect in high dimensions.\footnote{Some
of the above methods assume that the propensity scores can be consistently estimated using a sparse logistic model,
while others allow for the use of more flexible modeling strategies following, e.g.,
\citet{mccaffrey2004propensity}, \citet{van2007super}, or \citet{westreich2010propensity}.}

In this paper, we show that in settings where we are willing to entertain a sparse, well-specified linear model on the outcomes,
efficient inference of average treatment effects in high-dimensions is possible under more general assumptions than suggested by the
literature discussed above. Given linearity assumptions, we show that it is not necessary to consistently estimate treatment
propensities; rather, it is enough to rely on de-biasing techniques building on recent developments in the
high-dimensional inference literature
 \citep{javanmard2014confidence,javanmard2015biasing,van2014asymptotically,zhang2014confidence}.
In particular, in sparse linear models, we show that $\sqrt{n}$-consistent inference of average
treatment effects is possible provided we simply
require overlap, i.e., that the propensity score be uniformly bounded away from 0 and 1
for all values in the support of the pretreatment variables. We do not need to assume the
existence of a consistent estimator of the propensity scores, or any form of sparsity on the propensity model.

The starting point behind both our method and the doubly robust methods of \citet{belloni2013program},
\citet{chernozhukov2016double}, \citet{farrell2015robust}, \citet{van2011targeted}, etc., is a recognition
that high dimensional regression adjustments (such as the lasso) always shrink estimated effects, and that ignoring
this shrinkage may result in prohibitively biased treatment effect estimates. The papers on doubly robust
estimation then proceed to show that propensity-based adjustments can be used to compensate for this bias,
provided we have a consistent propensity model that converges fast enough to the truth. Conceptually, this
work builds on the result of \citet{rosenbaum1983central}, who showed that controlling for the propensity score
is sufficient to remove all biases associated with observed covariates, regardless of their functional form.

If we are willing to focus on high-dimensional linear models, however, it is possible to tighten the connection between
the estimation strategy and the objective of estimating the average treatment effect and, in doing so, extend the
number of settings where $\sqrt{n}$-consistent inference is possible. The key insight is that, in a linear model,
propensity-based methods are attempting to solve a needlessly difficult task when they seek to eliminate biases
of any functional form. Rather, in linear models, it is enough to correct for linear biases. In high dimensions, this can
still be challenging; however, we find that it is possible to approximately correct for such biases whenever we
assume overlap.

Concretely, we study the following two stage {approximate residual balancing} algorithm. 
First, we fit a regularized linear model for the outcome given the features separately in the two treatment groups.
In the current paper we focus on the elastic net \citep{zou2005regularization} and the
lasso \citep{chen1998atomic,tibshirani1996regression} for this component, and present formal results for the latter.
In a second stage, we re-weight the first stage residuals using weights that approximately balance all the features
between the treatment and control groups. Here we follow \citet{zubizarreta2015stable}, and optimize the implied balance
and variance provided by the weights, rather than the fit of the propensity score. Approximate balancing on all
pretreatment variables (rather than exact balance on a subset of features, as in a regularized regression,
or weighting using a regularized propensity model that may not be able to capture all relevant dimensions)
allows us to guarantee that the bias arising from a potential failure to adjust for a large
number of weak confounders can be bounded.
Formally, this second step of re-weighting residuals using the weights proposed by \citet{zubizarreta2015stable}
is closely related to de-biasing corrections studied in the high-dimensional regression literature
\citep{javanmard2014confidence,javanmard2015biasing,van2014asymptotically,zhang2014confidence};
we comment further on this connection in Section \ref{sec:theory}.

This approach also bears a close conceptual connection to work by \citet{chan2015globally},
\citet{deville1992calibration}, \citet{graham1, graham2},
\citet{hainmueller}, \citet{hellerstein1999imposing}, \citet{imai2014covariate} and \citet{zhao2016covariate},
who fit propensity models to the data under a constraint
that the resulting inverse-propensity weights exactly balance the covariate distributions between the treatment and
control groups, and find that these methods out-perform propensity-based methods that do not impose balance. Such
an approach, however, is only possible in low dimensions; in high dimensions where there are more covariates
than samples, achieving exact balance is in general impossible. One of the key findings of this paper is that, in high dimensions,
it is still often possible to achieve approximate balance under reasonable assumptions and that---when combined with
a lasso regression adjustment---approximate balance suffices for eliminating bias due to regularization.

In our simulations, we find that three features of the algorithm are important: $(i)$ the direct covariance adjustment based on the outcome data with regularization to deal with the large number of features, $(ii)$ the weighting using the relation between the treatment and the features, and $(iii)$ the fact that the weights are based on direct measures of imbalance rather than on estimates of the propensity score.
The finding that both weighting and regression adjustment are important is similar to conclusions drawn from the earlier literature on doubly robust estimation \citep[e.g.,][]{robins1997toward}, where combining both techniques was shown to extend the set of problems where efficient treatment effect estimation is possible.
The finding that weights designed to achieve balance perform better than weights based on the propensity score is consistent with findings in \citet{chan2015globally}, \citet{graham1, graham2}, \citet{hainmueller}, \citet{imai2014covariate}, and \citet{zubizarreta2015stable}.

Our paper is structured as follows. First, in Section \ref{sec:main}, we motivate our two-stage procedure using a simple bound for its estimation error. Then, in Section \ref{sec:theory}, we provide a formal analysis of our procedure under high-dimensional asymptotics, and we identify conditions under which approximate residual balancing is asymptotically Gaussian and allows for practical inference about the average treatment effect with dimension-free rates of convergence. Finally, in Section \ref{sec:simu}, we conduct a simulation experiment, and find our method to perform well in a wide variety of settings relative to other proposals in the literature.
A software implementation for \texttt{R} is available at \url{https://github.com/swager/balanceHD}.

\section{Estimating Average Treatment Effects in High Dimensions}
\label{sec:main}

\subsection{Setting and Notation}
\label{sec:setting}

Our goal is to estimate average treatment effects in the potential outcomes framework, or Rubin Causal Model \citep{rubin1974estimating,imbens2015causal}.
For each unit in a large population there is pair of (scalar) potential outcomes, $(\yin, \, \yie)$.  Each unit is assigned to the treatment or not, with the  treatment indicator denoted by $W_i\in\{0, \, 1\}$. Each unit is also characterized by a vector of covariates or features $X_i\in\mmr^p$, with $p$ potentially large, possibly larger than the sample size. For a random sample of size $n$ from this population, we observe the triple $(X_i, \, W_i, \, Y^\obs_i)$ for $i=1, \, \ldots, \, n$, where
\begin{equation}
Y^\obs_i=Y_i(W_i)=
\begin{cases}
\yie \hskip1cm & {\rm if}\  W_i=1,\\
\yin & {\rm if}\  W_i=0,
\end{cases}
\end{equation}
is the realized outcome, equal to the potential outcome corresponding to the actual treatment received. 
The total number of treated units is equal to $\nt$ and the number of control units equals $\nc$. 
We  frequently use the short-hand $\bx_\ccc$ and $\bx_\ttt$ for the feature matrices corresponding only to control or treated units respectively.
We write the propensity score, i.e., the conditional probability of receiving the treatment given features, as
$e(x)=\mathbb{P}[W_i=1|X_i=x]$  \citep{rosenbaum1983central}.
We focus  primarily on 
the conditional average treatment effect for the treated {sample},
\begin{equation}
\label{eq:taus}
\tau = \frac{1}{\nt} \sum_{\cb{i : W_i = 1}} \EE{\yin - \yie \cond X_i}.
\end{equation}
We note that the average treatment effect for the controls and the overall average effect can be handled similarly.
Throughout the paper we assume {unconfoundedness}, i.e., that conditional on the pretreatment variables,
treatment assignment is as good as random \citep{rosenbaum1983central};
we also assume a linear model for the potential  outcomes in both groups.

\begin{assu}[Unconfoundedness]
\label{assu:unconf}
$W_i\ \indep\  \p{\yin, \, \yie} \, \cond \, \ X_i$.
\end{assu}

\begin{assu}[Linearity]
\label{assu:linearity}
The conditional response functions satisfy 
$\mucc(x) = \EE{Y_i(0) \cond X = x} = x \cdot \betacc$ and
$\mutt(x) = \EE{Y_i(1) \cond X = x} = x \cdot \betatt$, for all $x \in \RR^p$. 
\end{assu}
Here, we will only use the linear model for the control outcome because we focus on the average effect for the treated units, but if we were interested in the overall average effect we would need linearity in both groups. The linearity assumption is strong,
but in high dimensions some strong structural assumption is in general needed for inference to be possible.
Then, given linearity, we have
\begin{equation}
\tau = \mutt - \mucc, \ \where \ \mutt =\hatx\cdot \betatt,\ \  
\mucc =\hatx\cdot \betacc,
 \ \eqand \ \hatx = \frac{1}{\nt} \sum_{i = 1}^n \mathbf{1}\p{\cb{W_i = 1}} X_i.
\end{equation}
Estimating the first term is easy: $\hmutt =\haty= \sum_{\cb{i : W_i = 1}} Y_i^\obs \, /\nt$ is unbiased for $\mutt$.
In contrast, estimating $\mucc$ is a major challenge, especially in settings where $p$ is large, and it is the main focus of the paper.

\subsection{Baselines and Background}
\label{sec:baselines}

We begin by reviewing two classical approaches to estimating $\mucc$, and thus also $\tau$, in the above linear model. 
The first is a weighting-based approach, which seeks to re-weight the control sample to make it look more like the treatment sample;
the second is a regression-based approach, which seeks to adjust for differences in features between treated and control units  by fitting an accurate model to the outcomes.
Though neither approach alone performs well in a high-dimensional setting with a generic propensity score,
we find that these two approaches can be fruitfully combined to obtain better estimators for $\tau$.

\subsubsection{Weighted Estimation}
\label{sec:weighted_avg}

A first approach is to re-weight the control dataset using weights $\gamma_i$ to make the weighted covariate distribution mimic the covariate distribution in the treatment population. Given the weights we estimate $\hmucc$ as a weighted average
\smash{$\hmucc = \sum_{\cb{i : W_i = 0}} \gamma_i \, Y_i^\obs$}.
The  standard way of selecting weights $\gamma_i$ uses the propensity score:
\smash{$\gamma_i = {e(X_i)}/(1 -  e(X_i))
\, / \, (\sum_{\cb{i : W_j = 0}} { e(X_j)}/(1 -  e(X_j))).$}
To implement these methods researchers typically substitute an estimate of the propensity score
into this expression.
Such {inverse-propensity weights} with non-parametric propensity score estimates have desirable asymptotic properties in settings with a small number of covariates \citep{hirano2003efficient}.
The finite-sample performance of methods based on inverse-propensity weighting can be poor, however,
both in settings with limited overlap in covariate distributions and in settings with many covariates.
A key difficulty is that estimating the treatment effect then involves dividing by $1 - \he(X_i)$, 
and so small inaccuracies in $\he(X_i)$ can have large effects, especially when $e(x)$ can be close to one;
this problem is often quite severe in high dimensions.

As discussed in the introduction, if the control potential outcomes $Y_i(0)$ have a linear dependence on
$X_i$, then using weights $\gamma_i$ that explicitly seek to balance the features $X_i$ is often advantageous \citep{deville1992calibration, chan2015globally, graham1, graham2, hainmueller, hellerstein1999imposing, imai2014covariate, zhao2016covariate, zubizarreta2015stable}.  This is a subtle but important improvement.
The motivation behind this approach is that, in a linear model, the bias for estimators based on weighted averaging depends solely on \smash{$\hatx - \sum_{\cb{i : W_i = 0}} \gamma_i \, X_i$}. Therefore getting the propensity model exactly right is less important than accurately matching the moments of \smash{$\hatx$}.
In high dimensions, however, exact balancing weights do not in general exist. When $p \gg \nc$, there will in general be no weights $\gamma_i$ for which $\hatx - \sum_{\cb{i : W_i = 0}} \gamma_i \, X_i = 0$, and even in settings where $p <\nc$ but $p$ is large such estimators would not have good properties. \citet{zubizarreta2015stable} extends the balancing weights approach to allow for weights that achieve approximate balance instead of exact balance; however, directly using his approach does not allow for $\sqrt{n}$-consistent estimation in a regime where $p$ is much larger than $n$.

\subsubsection{Regression Adjustments}

 A second approach is to  compute an estimator $\hbetacc$ for $\betacc$ using the $\nc$ control observations, and then estimate $\mucc$ as
$\hmucc = \hatx \cdot \hbetacc$.
In a low-dimensional regime with $p \ll \nc$,  the ordinary least squares estimator for $\betacc$
is a natural choice, and yields an accurate and unbiased estimate of $\mucc$. In high dimensions, however, the problem is more delicate: Accurate unbiased estimation of the regression adjustment is in general impossible, and methods such as the lasso, ridge regression, or the elastic net
may perform poorly when plugged in for $\betacc$, in particular when $\hatx$ is far away from $\overline{X}_{\rm c}$.
As stressed by \citet{belloni2014inference}, the problem with plain lasso regression adjustments is that features with a substantial difference in average values between the two treatment arms can generate large biases even if the coefficients on these features in the outcome regression are small.
Thus, a regularized regression that has been tuned to optimize goodness of fit on the outcome model is not appropriate whenever  bias in the treatment effect estimate due to failing to control for potential confounders is of concern.
To address this problem, \citet{belloni2014inference} propose running least squares regression on the union of two sets of selected variables, one selected by a lasso regressing the outcome on the covariates, and the other selected by a lasso logistic regression for the treatment assignment.
We note that estimating $\mucc$ by a regression adjustment $\hmucc=\hatx \cdot \hbetacc$, with $\hbetacc$ estimated by ordinary least squares on selected variables, is implicitly equivalent to using a weighted averaging estimator with weights $\gamma$ chosen to balance the selected features \citep{robins2007comment}.
The \citet{belloni2014inference} approach works well in settings where both the outcome regression
and the treatment regression are at least approximately sparse. However, when the propensity
is not sparse, we find that the performance of such double-selection methods is often poor. 

\subsection{Approximate Residual Balancing}
\label{sec:arb}

Here we propose a new method combining weighting and regression adjustments to overcome the limitations of each method.
In the first step of our method, we use a regularized linear model, e.g., the lasso or the elastic net, to obtain a pilot estimate of the treatment effect.   In the second step, we do ``approximate balancing'' of the regression residuals to estimate treatment effects:
We weight the residuals using weights that achieve approximate balance of the covariate distribution between treatment and control groups.
This step compensates for the potential bias of the pilot estimator that arises due to confounders that may be weakly correlated with the outcome but are important due to their correlation with
the treatment assignment.
We find that the regression adjustment is effective at capturing strong effects; the weighting on the other hand is effective at capturing small effects. The combination leads to an effective and simple-to-implement estimator for average treatment effects with many features.

We focus on a meta-algorithm that first computes an estimate $\hbetacc$ of $\betacc$ using the full sample of control units. This estimator may take a variety of forms, but typically it will involve some form of regularization to deal with the number of features. Second we compute weights $\gamma_i$ that balance the covariatees at least approximately, and apply these weights to the residuals \citep{cassel1976some,robins1994estimation}:
\begin{equation}
\label{eq:meta_alg}
\hmucc =  \hatx \cdot \hbetacc + \sum_{\cb{i : W_i = 0}} \gamma_i \, \p{Y_i^\obs - X_i \cdot \hbetacc}  . 
\end{equation}
In other words, we fit a model parametrized by $\betacc$ to capture some of the strong signals, and then use direct numerical re-balancing of the control data on the features to extract left-over signal from the residuals $Y_i^\obs - X_i \cdot \hbetacc$.
Ideally, we would hope for the first term to take care of any strong effects, while the re-balancing of the residuals can efficiently take care of the small spread-out effects.
Our theory and experiments will verify that this is in fact the case.

A major advantage of the functional form in \eqref{eq:meta_alg} is that it yields a simple and powerful theoretical guarantee, as stated below. Recall that $\bxcc$ is the feature matrix for the control units. 
Consider the  difference between $\hmucc$ and $\mucc$ for our proposed approach:
$ \hmucc - \mucc= (\hatx - \bxcc^\top \gamma)  \cdot (\hbetacc - \betacc)+\gamma \cdot \varepsilon$,
where $\varepsilon$ is the intrinsic noise $\varepsilon_i =  Y_i(0) - X_i \cdot \betacc$.
With only the regression adjustment and no weighting, the difference would be
$\hat{\mu}_{{\rm c},{\rm reg}} - \mucc= (\hatx - \overline{X}_{\rm c}) \cdot (\hbetacc - \betacc) + \mathbf{1} \cdot \varepsilon/\nc$,
and with only the weighting the difference would be
$\hat{\mu}_{{\rm c},{\rm weight}} - \mucc=(\hatx - \bxcc^\top \gamma) \cdot \betacc + \gamma \cdot \varepsilon$.
Without any adjustment, just using the average outcome for the controls as an estimator for $\mucc$, the difference between the estimator for $\mucc$ and its actual value would be
$\hat{\mu}_{{\rm c},{\rm no-adj}} - \mucc=(\hatx - \overline{X}_{\rm c}) \cdot \betacc+\mathbf{1} \cdot \varepsilon/\nc$.
The regression reduces the bias from $(\hatx - \overline{X}_{\rm c}) \cdot \betacc$ to $(\hatx - \overline{X}_{\rm c}) \cdot (\hbetacc - \betacc)$, which will be substantial reduction if the estimation error $(\hbetacc - \betacc)$ is small relative to $\betacc$. The weighting further reduces this to $(\hatx - \bxcc^\top \gamma) \cdot (\hbetacc - \betacc)$, which may be helpful if there is a substantial difference between $\hatx$ and $\overline{X}_{\rm c}$.
This result, formalized below, shows the complimentary nature of the regression adjustment and the weighting.
All proofs are given in the appendix.

\begin{prop}
\label{prop:main}
The estimator \eqref{eq:meta_alg} satisfies
$\abs{\hmucc - \mucc} \leq \Norm{\hatx - \bxcc^\top \gamma}_\infty \Norm{\hbetacc - \betacc}_1 + \abs{\sum_{\cb{i : W_i = 0}} \gamma_i \, \varepsilon_i}$.
\end{prop}

This result decomposes the error of $\hmuC$ into two parts. The first is a bias term reflecting the dimension $p$ of the covariates; the second term is a {variance term} that does not depend on it. The upshot is that the bias term, which encodes the high-dimensional nature of the problem, involves a product of two factors that can both other be made reasonably small; we will focus on regimes where the first term should be expected to scale as $\oo(\sqrt{\log(p)/n})$, while the second term scales as $\oo(k\sqrt{\log(p)/n})$ where $k$ is the sparsity of the outcome model.
If we are in a sparse enough regime (i.e., $k$ is small enough), Proposition \ref{prop:main} implies that our procedure
will be variance dominated; and, under these conditions, we also show that it is $\sqrt{n}$-consistent.

In order to exploit Proposition \ref{prop:main}, we need to make concrete choices for the
weights $\gamma$ and the parameter estimates $\hbetacc$.
First, just like \citet{zubizarreta2015stable}, we choose our weights $\gamma$ to directly optimize
the bias and variance terms in Proposition \ref{prop:main}; the functional form of $\gamma$ is given in
\eqref{eq:gamma_def}, where $\zeta \in (0, \, 1)$ is a tuning parameter.
We refer to them as \emph{approximately balancing weights} since they seek to make the mean of the re-weighted control sample, namely $\bxcc^\top \gamma$, match the treated sample mean $\hatx$ as closely as possible. 
The positivity constraint on $\gamma_i$ in \eqref{eq:gamma_def} aims to prevent the method from extrapolating too aggressively,
while the upper bound is added for technical reasons discussed in Section \ref{sec:theory}.
Meanwhile, for estimating $\hbetacc$, we simply need to use an estimator that achieves
good enough risk bounds under $L_1$-risk. In our analysis, we focus on the
lasso \citep{chen1998atomic,tibshirani1996regression}; however, in experiments, we use
the elastic net for additional stability \citep{zou2005regularization}.
Our complete algorithm is described in Procedure \ref{alg:approx_balance}.

\begin{algbox}[t]
\myalg{alg:approx_balance}{Approximately Residual Balancing with Elastic Net}{
The following algorithm estimates the average treatment effect on the treated by approximately balanced residual weighting.
Here, $\zeta \in (0, \, 1)$, $\alpha \in (0, \, 1]$ and $\lambda > 0$ are tuning parameters.
This procedure is implemented in our \texttt{R} package \texttt{balanceHD}; we default to $\zeta = 0.5$ and $\alpha = 0.9$, and select $\lambda$ by cross-validation using the \texttt{lambda.1se} rule from the \texttt{glmnet} package \citep{friedman2010regularization}.
The optimization problem \eqref{eq:gamma_def} is a quadratic program, and so can be solved using
off-the-shelf convex optimization software; we use the interior point solver \texttt{mosek} by default
\citep{mosek}.
\begin{enumerate}
\item Compute positive approximately balancing weights $\gamma$ as
\begin{align}
\label{eq:gamma_def}
\gamma = \argmin_{\tgamma}\cb{ \p{1 - \zeta} \Norm{\tgamma}_2^2 + \zeta \Norm{\hatx - \bxcc^\top \tgamma}_\infty^2 \text{ s.~t. } \!\!\!\!\!\! \sum_{\cb{i : W_i = 0}} \tgamma_i = 1 \eqand 0 \leq \tgamma_i \leq n_{\rm c}^{-2/3}}.
\end{align}
\item Fit $\betacc$ in the linear model
using a lasso or an elastic net,
\begin{equation}
\label{eq:elnet}
 \hbetacc = \argmin_\beta \cb{\sum_{\cb{i : W_i = 0}} \p{Y_i^\obs - X_i \cdot \beta}^2
 + \lambda \p{\p{1 - \alpha} \Norm{\beta}_2^2 + \alpha \Norm{\beta}_1}}.
\end{equation}
\item Estimate the average treatment effect $\tau$ as 
\begin{equation}
\htau=\haty- \left( \hatx \cdot \hbetacc + \sum_{\cb{i : W_i = 0}}\gamma_i \, \p{Y_i^\obs - X_i \cdot \hbetacc}\right).
\end{equation}
\end{enumerate}
}
\end{algbox}

Finally, although we do use this estimator in the present paper, we note that an analogous estimator
for the average treatment effect $\EE{Y(1) - Y(0)}$ can also be constructed:
\begin{equation}
\label{eq:gamma_def_ate}
\begin{split}
&\htau_{ATE} = \bX\p{\hbetatt - \hbetacc} 
+\sum_{\cb{i : W_i = 1}} \gamma_{{\rm t},i} \, \p{Y_i^\obs - X_i \cdot \hbetatt}
-\sum_{\cb{i : W_i = 0}} \gamma_{{\rm c},i} \, \p{Y_i^\obs - X_i \cdot \hbetacc}, \where \\
&\gamma_{\rm t} = \argmin_{\tgamma}\cb{ \p{1 - \zeta} \Norm{\tgamma}_2^2 + \zeta \Norm{\bX - \bxtt^\top \tgamma}_\infty^2 \text{ subject to } \!\!\!\!\!\! \sum_{\cb{i : W_i = 1}} \tgamma_i = 1 \eqand 0 \leq \tgamma_i \leq n_{\rm t}^{-2/3}},
\end{split}
\end{equation}
and $\gamma_{\rm c}$ is constructed similarly. This method can be analyzed using the same
tools developed in this paper, and is available in our software package \texttt{balanceHD}.
The conditions required for $\sqrt{n}$-consistent inference of $\tau_{ATE}$ using \eqref{eq:gamma_def_ate}
directly mirror the conditions (sparsity, overlap, etc.) listed in Section \ref{sec:theory} for inference
about the average treatment effect on the treated.

\subsection{Connection to Doubly Robust Estimation}

The idea of combining weighted and regression-based approaches to treatment effect estimation
has a long history in the causal inference literature. Given estimated propensity scores \smash{$\he(X_i)$},
\citet{cassel1976some} and \citet{robins1994estimation} propose using an augmented inverse-propensity
weighted (AIPW) estimator,
\begin{equation}
\label{eq:aipw}
\hmucc^{(AIPW)} = \hatx \cdot \hbetacc + \sum_{\cb{i : W_i = 0}} \frac{\he(X_i)}{1 - \he(X_i)} \, \p{Y_i^\obs - X_i \cdot \hbetacc} \, \bigg / \, \sum_{\cb{i : W_i = 0}} \frac{\he(X_i)}{1 - \he(X_i)};
\end{equation}
the difference between this estimator and ours is that they obtain their weights $\gamma_i$ for \eqref{eq:meta_alg}
via propensity-score modeling instead of quadratic programming. Estimators of this type have several desirable aspects:
they are ``doubly robust'' in the sense that they are consistent whenever either the propensity fit $\he(\cdot)$
or the outcome fit $\hbetacc$ is consistent, and they are asymptotically efficient in a semiparametric specification
\citep{hahn1998role,hirano2003efficient,robins1,tsiatis2007semiparametric}.
However, a practical concern with this class of methods is that they may perform less well when \smash{$1 - \he(X_i)$}
is close to 0 \citep{hirano2003efficient, schafer}.
Several higher-order refinements to the simple AIPW estimator \eqref{eq:aipw} have also been considered in the literature.
In particular, \citet{schafer} use the inverse-propensity weights  \smash{$\he(X_i)/(1 - \he(X_i))$}
as sample weights when estimating $\hbetacc$, while \citet{scharfstein1999adjusting} and \citet{van2006targeted}
consider adding these weights as features in the outcome model; see also \citet{robins2007comment} and \citet{tan2010bounded} for further
discussion.

\citet{belloni2013program} and \citet{farrell2015robust}
study the behavior of AIPW in high dimensions, and establish conditions under which they can reach efficiency
when both the propensity function and the outcome model are consistently estimable.
Intriguingly, in low dimensions, doubly robust methods are not necessary for achieving semiparametric efficiency.
This rate can be achieved by either non-parametric inverse-propensity weighting or non-parametric regression adjustments on their own \citep{chen2008semiparametric,hirano2003efficient};
doubly robust methods can then be used to relax the regularity conditions
needed for efficiency \citep{farrell2015robust,robins2008higher}.
Conversely, in high-dimensions, both weighting and regression adjustments are required for $\sqrt{n}$-consistency
\citep{belloni2013program,farrell2015robust,robins1997toward}.

Although our estimator \eqref{eq:meta_alg} is cosmetically quite closely related to the AIPW estimator \eqref{eq:aipw},
the motivation behind it is quite different. A common description of the AIPW estimator is that it tries to estimate two
different nuisance components, i.e., the outcome model $\hmucc$ and the propensity model $\he$; it then achieves consistency
if either of these components is itself estimated consistently, and efficiency if both components are estimated at fast enough
rates. In contrast, our approximate residual balancing estimator bets on linearity twice: once in fitting the
outcome model via the lasso, and once in de-biasing the lasso via balancing weights \eqref{eq:gamma_def}.

By relying more heavily on linearity, we can considerably extend the set of problem under which $\sqrt{n}$-consistent
is possible (assuming linearity in fact holds). As a concrete example, a simple analysis of AIPW estimation in high-dimensional
linear models would start by assuming that the lasso is $o_P(n^{-1/4})$ consistent in root-mean squared error,\footnote{A more
careful analysis of AIPW estimators can trade off the accuracy of the propensity and main effect models and, instead
of requiring that both the propensity and outcome models can be estimated at $o_P(n^{-1/4})$ rates, only
assumes that the product of the two rates be bounded as $o_P(n^{-1/2})$; see, e.g., \citet{farrell2015robust}.
In high dimensions, this amounts to assuming that the outcome and propensity models are both well specified and
sparse, with respective sparsity levels $k_\beta$ and $k_e$ satisfying $k_\beta k_e \ll n / \log(p)^2$.
AIPW can thus be preferable to ARB given sparse enough and well specified propensity models, with $k_e \ll \sqrt{n}/\log(p)$.}
which can be attained via the lasso assuming a $k$-sparse true model with sparsity level $k \ll \sqrt{n}/\log(p)$; and this is, in fact,
exactly the same condition we assume in Theorem \ref{theo:overlap}. Then, in addition to this requirement on the outcome model,
AIPW estimators still need to posit the existence of an $o_P(n^{-1/4})$ consistent estimator of the treatment propensities,
whereas we do not need to assume anything about the treatment assignment mechanism beyond overlap.
The reason for this phenomenon is that the task of balancing (which is all that is needed to correct for the bias
of the lasso in a linear model) is different from the task of estimating the propensity score---and is in fact often substantially
easier.\footnote{The above distinctions are framed are in a situation where the statistician
starts with a set of high-dimensional covariates, and needs to find a way to control for all of them at
once. In this setting, linearity is a strong assumption, and so it is not surprising that making this assumption
lets us considerably weaken requirements on other aspects of the problem.
In other applications, however, the statistician may have started with low-dimensional data, but then created
a high-dimensional design by listing series expansions of the original data, interactions, etc. In this setting, 
linearity is replaced with smoothness assumptions on the outcome model (since any smooth function can be well approximated using a
large enough number of terms from an appropriately chosen series expansion). Here, variants of our
procedure can be more directly compared with doubly robust methods and, in particular, the
$\gamma_i$ in fact converge to the oracle inverse-propensity weights $e(X_i)/(1 - e(X_i))$;
see \citet{hirshberg2017balancing} and \citet{wang2017approximate} for a discussion and further results.}

\subsection{Related Work}

Our approximately balancing weights \eqref{eq:gamma_def} are inspired by the recent litearture on balancing
weights \citep{chan2015globally, deville1992calibration, graham1, graham2, hainmueller, hellerstein1999imposing,
hirr, imai2014covariate, zhao2016covariate}.
Most closely related, \citet{zubizarreta2015stable} proposes estimating $\tau$ using the re-weighting
formula as in Section \ref{sec:weighted_avg} with weights
\begin{align}
\label{eq:gamma_def_zubi}
\gamma = \argmin_{\tgamma} \cb{ \Norm{\tgamma}_2^2 \text{ subject to } \sum \tgamma_i = 1,\ \tgamma_i\geq 0,\ 
\Norm{\hatx - \bxcc^\top \tgamma}_\infty \leq t
},
\end{align}
where the tuning parameter is $t$; he calls these weights \emph{stable balancing weights}. These weights
are of course equivalent to ours, the only difference being that Zubizarreta bounds imbalance in constraint
form whereas we do so in Lagrange form.
The main conceptual difference between our setting and that of \citet{zubizarreta2015stable} is that he considers
problem settings where $p < \nc$, and then considers $t$ to be a practically small tuning parameter,
e.g., $t = 0.1\sigma$ or $t = 0.001 \sigma$. However, in high dimensions, the optimization
problem \eqref{eq:gamma_def_zubi} will not in general be feasible for small values of $t$; and in fact
the bias term \smash{$\Norm{\hatx - \bxcc^\top \gamma}_\infty$} becomes the dominant source of error
in estimating $\tau$. We call our weights $\gamma$ ``approximately'' balancing in order to remind the reader of this fact.
In settings where it is only possible to achieve approximate balance, weighting alone as
considered in \citet{zubizarreta2015stable} will not yield a $\sqrt{n}$-consistent estimate
of the average treatment effect, and it is necessary to use a regularized regression adjustment as
in \eqref{eq:meta_alg}.

Similar estimators have been considered by \citet{graham1, graham2} and \citet{hainmueller} in a setting where exact balancing is possible, with slightly different objection functions. For example, \citet{hainmueller} uses  $\sum_{i} \gamma_i\log(\gamma_i)$ instead of $\sum_{i} \gamma_i^2$, leading to
\begin{align}
\label{eq:gamma_def_hainmuller}
\gamma = \argmin_{\tgamma} \cb{\sum_{\cb{i : W_i = 0}} \tgamma_i\log\p{\tgamma_i} \text{ subject to } \sum \tgamma_i = 1,\ \tgamma_i\geq 0,\ \ 
\Norm{\hatx - \bxcc^\top \tgamma}_\infty=0}.
\end{align}
This estimator has attractive conceptual connections to logistic regression and maximum entropy estimation, and in a low dimensional setting where $W|X$ admits a well-specified logistic model the methods of \citet{graham1,graham2} and \citet{hainmueller} are doubly robust 
\citep{zhao2017entropy}; see also \citet{hirr}, \citet{imbensel}, and \citet{newey2004higher}.
In terms of our immediate concerns, however, the variance of $\htau$ depends on $\gamma$ through \smash{$\Norm{\gamma}_2^2$} and not \smash{$\sum \gamma_i\log\p{\gamma_i}$}, so our approximately balancing weights are more directly induced by our statistical objective than those defined in \eqref{eq:gamma_def_hainmuller}.

Finally, in this paper, we have emphasized an asymptotic analysis point of view, where we evaluate estimators
via their large sample accuracy. From this perspective, our estimator---which combines weighting with a regression
adjustment as in \eqref{eq:meta_alg}---appears to largely dominate pure weighting estimators; in particular, in
high dimensions, we achieve $\sqrt{n}$-consistency whereas pure weighting estimators do not. On the other hand,
stressing practical concerns, \citet{rubin2008objective} strongly argues that ``designed based'' inference leads
to more credible conclusions in applications by better approximating randomized experiments. In our context, design based inference
amounts to using a pure weighting estimator of the form $\sum \gamma_i Y_i$ where the $\gamma_i$ are chosen
without looking at the $Y_i$. The methods considered by \citet{chan2015globally}, \citet{graham1}, \citet{hainmueller},
\citet{zubizarreta2015stable}, etc., all fit within this design-based paradigm, whereas ours does not.

\section{Asymptotics of Approximate Residual Balancing}
\label{sec:theory}

\subsection{Approximate Residual Balancing as Debiased Linear Estimation}

As we have already emphasized, approximate residual balancing is a method that enables us
to makes inferences about average treatment effects without needing to estimate treatment propensities as
nuisance parameters; rather, we build on recent developments on inference in high-dimensional linear models
 \citep{cai2015confidence,javanmard2014confidence,javanmard2015biasing,
 ning2014general,van2014asymptotically,zhang2014confidence}.
Our main goal is to understand the asymptotics our estimates for $\mucc = \hatx \cdot \betacc$.
In the interest of generality, however, we begin by considering a broader problem, namely that
of estimating generic contrasts $\xi \cdot \betacc$ in high-dimensional linear models.
This detour via linear theory will help highlight the statistical phenomena that make approximate
residual balancing work, and explain why---unlike the methods of \citet{belloni2013program},
\citet{chernozhukov2016double} or \citet{farrell2015robust}---our method does not require
consistent estimability of the treatment propensity function $e(x)$.

The problem of estimating \emph{sparse} linear contrasts $\xi \cdot \betacc$ in high-dimensional regression
problems has received considerable attention, including notable recent contributions by
\citet{javanmard2014confidence,javanmard2015biasing}, \citet{van2014asymptotically}, and \citet{zhang2014confidence}.
These papers, however, exclusively consider the setting where $\xi$ is a sparse vector,
and, in particular, focus on the case where $\xi$ is the $j$-th basis vector $e_j$, i.e.,
the target estimand is the $j$-th coordinate of $\betacc$.
But, in our setting, the contrast vector $\hatx$ defining our estimand
$\mucc = \hatx \cdot \betacc$ is random and thus generically dense;
moreover, we are interested in applications where $m_{\rm t} = \mathbb{E}[\hatx]$ itself may also be dense.
Thus, a direct application of these method is not appropriate in our problem.\footnote{As
a concrete example, Theorem 6 of \citet{javanmard2014confidence} shows that their
debiased estimator \smash{$\hbeta_{\text{c}}^{(\text{debiased})}$} satisfies
\smash{$ \sqrt{n_{\rm c}}(\hbeta_{\text{c}}^{(\text{debiased})} - \betacc) = Z + \Delta$},
where $Z$ is a Gaussian random variable with desirable properties and
\smash{$\lVert\Delta\rVert_\infty = o(1)$}. If we simply consider sparse contrasts of
\smash{$\hbetacc$}, then this error term $\Delta$ is negligible; however, in our setting, we would have
a prohibitively large error term $\hatx \cdot \Delta$ that may grow polynomially in $p$.}

An extension of this line of work to the problem of estimating dense, generic contrasts
$\theta = \xi \cdot \betacc$ turns out to be closely related to our 
approximate residual balancing method for treatment effect estimation. To make this connection
explicit, define the following estimator:
\begin{align}
\label{eq:htheta}
&\htheta = \xi \cdot \hbetacc + \sum_{\cb{i : W_i = 0}} \gamma_i \p{Y_i^\obs - X_i \cdot \hbetacc}, \where \\
\label{eq:gamma_bdd}
&\gamma = \argmin_{\tgamma} \cb{\Norm{\tgamma}_2^2 \text{ subject to }  \Norm{\xi - \bxcc^\top \tgamma}_\infty \leq K \sqrt{\frac{\log(p)}{\nc}},  \ \max_i \abs{\tgamma_i} \leq \nc^{-2/3}},
\end{align}
$\hbetacc$ is a properly tuned sparse linear estimator, and $K$ is a tuning parameter discussed below.
If we set $\xi \leftarrow \hatx$, then this estimator is nothing but our treatment effect estimator
from Procedure \ref{alg:approx_balance}.\footnote{Here, we phrased the imbalance constraint in constraint
form rather than in Lagrange form; the reason for this is that, although there is a 1:1 mapping between these
two settings, we found the former easier to work with formally whereas the latter appears to yield more consistent
numerical performance. We also dropped the constraints $\sum \gamma_i = 1$ and $\gamma_i \geq 0$ for now,
but will revisit them in Section \ref{sec:direct}.}
Conversely, in the classical parameter estimation setting with $\xi \leftarrow e_j$, the above procedure is
algorithmically equivalent to the one proposed by \citet{javanmard2014confidence,javanmard2015biasing}.
Thus, the estimator \eqref{eq:htheta} can be thought of as an adaptation of the method of
\citet{javanmard2014confidence,javanmard2015biasing} that debiases \smash{$\hbetacc$}
specifically along the direction of interest $\xi$.

We begin our analysis in Section \ref{sec:xi} by considering a general version of \eqref{eq:htheta}
under fairly strong ``transformed independence design'' generative assumptions on $\bxcc$.
Although these assumptions may be too strong to be palatable in practical data analysis, this
result lets us make a crisp conceptual link between approximate residual balancing and the debiased
lasso. In particular, we find that (Theorem \ref{theo:debiased_prediction}), \smash{$\htheta$} from \eqref{eq:htheta} is $\sqrt{n}$-consistent
for $\theta$ provided \smash{$\xi^\top \Sigma_{\rm c}^{-1} \xi = \oo(1)$}, where $\Sigma_{\rm c}$ is the covariance of $\bxcc$.
Interestingly, if \smash{$\Sigma_{\rm c} = I_{p \times p}$}, then in general
\smash{$\xi^\top \Sigma_{\rm c}^{-1} \xi = \Norm{\xi}_2^2 = \oo(1)$}
if and only if $\xi$ is very sparse, and so the classical de-biased lasso theory reviewed above is
essentially sharp despite only considering the sparse-$\xi$ case \citep[see also][]{cai2015confidence}.
On the other hand, whenever $\Sigma_{\rm c}$ has latent correlation structure, it is possible to have
\smash{$\xi^\top \Sigma_{\rm c}^{-1} \xi = \oo(1)$} even when $\xi$ is dense and $\Norm{\xi}_2 \gg 1$,
provided that $\xi$ is aligned with the large latent components of $\Sigma_{\rm c}$.
We also note that, in the application to treatment effect estimation, \smash{$\hatx^\top \Sigma_{\rm c}^{-1} \hatx$}
will in general be much larger than 1; however, in Corollary \ref{coro:debiased_catt} we show how to
overcome this issue.

To our knowledge, this was the first result for $\sqrt{n}$-consistent inference about dense contrasts of $\betacc$
at the time we first circulated our manuscript. We note, however, simultaneous and independent work
by \citet{zhu2016linear}, who developed a promising method for testing hypotheses of the form
$\xi \cdot \betacc = 0$ for potentially dense vectors $\xi$; their approach uses an orthogonal moments
construction that relies on regressing $\xi \cdot X_i$ against a $p - 1$ dimensional design that captures
the components of $X_i$ orthogonal to $\xi$.

Finally, in Section \ref{sec:direct}, we revisit the specific problem of high-dimensional treatment
effect estimation via approximate residual balancing under substantially weaker assumptions on the
design matrix $\bxcc$: Rather than assuming a generative ``transformed independence design'' model,
we simply require overlap and standard regularity conditions. The cost of relaxing our assumptions on
$\bxcc$ is that we now get slightly looser performance guarantees; however, our asymptotic error
rates are still in line with those we could get from doubly robust methods.
We also discuss practical, heteroskedasticity-robust confidence intervals for $\tau$. Through our analysis, we assume that $\hbetacc$
is obtained via the lasso; however, we could just as well consider, e.g., the square-root lasso
\citep{belloni2011square}, sorted $L_1$-penalized regression \citep{bogdan2015slope,su2016slope},
or other methods with comparable $L_1$-risk bounds.

\subsection{Debiasing Dense Contrasts}
\label{sec:xi}

As we begin our analysis of $\htheta$ defined in \eqref{eq:htheta}, it is first important to
note that the optimization program \eqref{eq:gamma_bdd} is not always feasible.
For example suppose that $p = 2\nc$, that $\bxcc = (I_{\nc \times \nc} \ I_{\nc \times \nc})$, and
that $\xi$ consists of $n$ times ``$1$''  followed by $n$ times ``$-1$''; then
$\Norm{\xi - \bxcc^\top \gamma}_\infty \geq 1$ for any $\gamma \in \RR^{\nc}$,
and the approximation error does not improve as $\nc$ and $p$ both get large.
Thus, our first task is to identify a class of problems for which \eqref{eq:gamma_bdd}
has a solution with high probability. The following lemma establishes such a result for random
designs, in the case of vectors $\xi$ for which $\xi^\top \Sigma_{\rm c}^{-1} \xi$ is bounded; here
$\Sigma_{\rm c} = \Var{X_i \cond W_i = 0}$ denotes the population variance of control features.
We also rely on the following regularity condition, which will be needed for an application
of the Hanson-Wright concentration bound for quadratic forms following \citet{rudelson2013hanson}.

\begin{assu}[Transformed Independence Design]
\label{assu:indep}
Suppose that we have a sequence of random design problems with\footnote{In order
to simplify our exposition, this assumption implicitly rules out the use of an intercept.
Our analysis would go through verbatim, however, if we added an intercept $X_1 = 1$
to the design.}
\smash{$\bxcc = Q \, \Sigma_{\rm c}^{\frac{1}{2}}, \ \where \ \EE{Q_{ij}} = 0, \ \Var{Q_{ij}} = 1$},
for all indices $i$ and $j$, and the individual entries $Q_{ij}$ are all independent.
Moreover suppose that the $Q$-matrix is sub-Gaussian for some $\varsigma > 0$,
$\EE{\exp\sqb{t\p{Q_{ij} - \EE{Q_{ij}}}}} \leq \exp\sqb{\varsigma^2 t^2 / 2} \text{ for any } t > 0$,
and that $(\Sigma_{\rm c})_{jj} \leq S$ for all $j = 1, \, ... ,\, p$.
\end{assu}

\begin{lemm}
\label{lemm:OLS_balance}
Suppose that we have a sequence of problems for which Assumption \ref{assu:indep} holds
and, moreover, $\xi^\top \Sigma_{\rm c}^{-1} \xi \leq V$ for some constant $V > 0$. Then,
there is a universal constant $C > 0$ such that, setting $K = C\varsigma^2 \sqrt{VS}$, the optimization
problem \eqref{eq:gamma_bdd} is feasible with probability tending to 1; and, in particular, the
constraints are satisfied by
$\gamma^*_i = \frac{1}{\nc} \xi^\top \Sigma_{\rm c}^{-1} X_i$.
\end{lemm}

The above lemma is the key to our analysis of approximate residual balancing. Because, with high
probability, the weights $\gamma^*$ from Lemma \ref{lemm:OLS_balance} provide one feasible solution to
the constraint in \eqref{eq:gamma_bdd}, we conclude that, again with high probability, the actual
weights we use for approximate residual balancing must satisfy
$\Norm{\gamma}_2^2 \leq \Norm{\gamma^*}_2^2 \approx \nc^{-1} \xi^\top \Sigma_{\rm c}^{-1} \xi$.
In order to turn this insight into a formal result, we need assumptions on both the sparsity of
the signal and the covariance matrix $\Sigma_{\rm c}$.

\begin{assu}[Sparsity]
\label{assu:spar}
We have a sequence of problems indexed by $n$, $p$, and $k$ such that
the parameter vector $\betacc$ is $k$-sparse, i.e., $\Norm{\betacc}_0 \leq k$, and that
$k \, \log(p)/\sqrt{n} \rightarrow 0$.\footnote{In recent literature, there has been some
interest in methods that require only require approximate, rather than exact, $k$-sparsity.
We emphasize that our results also hold with approximate rather than exact sparsity, as
we only use our sparsity assumption to get bounds on \smash{$\lVert\hbetacc - \betacc\rVert_1$}
that can be used in conjunction with Proposition \ref{prop:main}. For simplicity of exposition, however,
we restrict our present discussion to the case of exact sparsity.}
\end{assu}

The above sparsity requirement is quite strong. However, many analyses that 
seek to establish asymptotic normality in high dimensions rely on such an assumption.
For example, \citet{javanmard2014confidence}, \citet{van2014asymptotically},
and \citet{zhang2014confidence} all make this assumption 
when seeking to provide confidence intervals for individual components of $\betacc$;
\citet{belloni2014inference} use a similar assumption where they allow for additional non-zero
components, but they assume that beyond the largest $k$ components with $k$ satisfying
the same sparsity condition, the remaining non-zero elements of $\betacc$ are sufficiently
small that they can be ignored, in what they refer to as approximate sparsity.\footnote{There
are, of course, some exceptions to this assumption.
In recent work, \citet{javanmard2015biasing} show that inference of $\betacc$ is possible even when
$k \ll n \, / \, \log(p)$ in a setting where $X$ is a random Gaussian matrix with either a
known or extremely sparse population precision matrix; \citet{wager2016high}
show that lasso regression adjustments allow for efficient average treatment effect estimation
in randomized trials even when $k \ll n \, / \, \log(p)$; while the method of \citet{zhu2016linear}
for estimating dense contrasts $\xi \cdot \betacc$
does not rely on sparsity of $\betacc$, and instead places assumptions on the joint distribution of
$\xi \cdot X_i$ and the individual regressors. The point in common between these
results is that they let us weaken the sparsity requirements at the expense of
strengthening our assumptions about the $X$-distribution.}

Next, our analysis builds on well-known bounds on the estimation error of the lasso \citep{bickel2009simultaneous,hastie2015statistical}
that require  $\bxcc$ to satisfy a form of the restricted eigenvalue condition.
Below, we make a restricted eigenvalue assumption on \smash{$\Sigma_{\rm c}^{1/2}$}; then, we will use results
from \citet{rudelson2013reconstruction} to verify that this also implies a restricted eigenvalue
condition on $\bxcc$.

\begin{assu}[Well-Conditioned Covariance]
\label{assu:re}
Given the sparsity level $k$ specified above, the covariance matrix $\Sigma_{\rm c}^{1/2}$ of the
control features satisfies the $\cb{k, \, 2\omega, \, 10}$-restricted eigenvalue defined
as follows, for some $\omega > 0$. For $1 \leq k \leq p$ and $L \geq 1$, define the set $\cset_k(L)$ as
\begin{equation}
\cset_k(L) = \cb{\beta \in \RR^p : \Norm{\beta}_1 \leq L \, \sum_{j = 1}^k \abs{\beta_{i_j}} \ \text{ for some }\ 1 \leq i_1 < ... < i_j \leq p}. 
\end{equation}
Then, $\Sigma_{\rm c}^{1/2}$ satisfies the $\cb{k, \, \omega, \, L}$-restricted eigenvalue condition if
$\beta^\top \Sigma_{\rm c} \beta \geq \omega  \Norm{\beta}_2^2 \, \text{ for all } \, \beta \, \in \, \cset_k(L)$.
\end{assu}

\begin{theo}
\label{theo:debiased_prediction}
Under the conditions of Lemma \ref{lemm:OLS_balance}, suppose that the control
outcomes $Y_i(0)$ are drawn from a sparse, linear model as in
Assumptions \ref{assu:unconf}, \ref{assu:linearity}, \ref{assu:indep} and \ref{assu:spar},
that \smash{$\Sigma_{\rm c}^{1/2}$} satisfies the restricted eigenvalue property (Assumption \ref{assu:re}),
and that we have a minimum estimand size\footnote{The minimum estimand size assumption is needed to
rule out pathological superefficient behavior. As a concrete example, suppose that
$X_i \sim \nn\p{0, \, I_{p \times p}}$, and that $\xi_j = 1/\sqrt{p}$ for $j = 1, \, ..., \, p$ with $p \gg \nc$. Then, with
high probability, the optimization problem \eqref{eq:gamma_bdd} will yield $\gamma = 0$. This leaves
us with a simple lasso estimator $\smash{\htheta = \xi \cdot \hbetacc}$ whose risk
scales as \smash{$\mathbb{E}[(\htheta - \theta)^2] = \oo(k^2\log(p)/(p \nc)) \ll 1/\nc$}. The problem
with this superefficient estimator is that it is not necessarily asymptotically Gaussian.}
 $\Norm{\xi}_\infty \geq \kappa > 0$.
Suppose, moreover, that we have homoskedastic noise:
\smash{$\operatorname{Var}[\varepsilon_i(0) \cond X_i] = \sigma^2$} for all $i = 1, \, ..., \, n$,
and also that the response noise \smash{$\varepsilon_i(0) := Y_i(0) - \mathbb{E}[Y_i(0) \cond X_i]$}
is uniformly sub-Gaussian with parameter $\upsilon^2 S > 0$.
Finally, suppose that we estimate \smash{$\htheta$} using \eqref{eq:htheta}, with the optimization
parameter $K$ selected as in Lemma \ref{lemm:OLS_balance} and the lasso penalty parameter set
to $\lambda_n = 5 \varsigma^2 \upsilon \sqrt{\log\p{p} / \nc}$. Then, $\htheta$ is asymptotically Gaussian,
\begin{equation}
\label{eq:debiased_prediction}
\p{\htheta - \theta}\,\big/\,{\Norm{\gamma}_2} \Rightarrow \nn\p{0, \, \sigma^2}, \ \  \nc \Norm{\gamma}_2^2 \,\big/\,  \xi^\top \Sigma_{\rm c}^{-1} \xi \leq 1 + o_p(1).
\end{equation}
\end{theo}

The statement of Theorem \ref{theo:debiased_prediction} highlights a connection between our
debiased estimator \eqref{eq:htheta}, and the ordinary least-squares (OLS) estimator. Under classical large-sample
asymptotics with $n \gg p$, it is well known that the OLS estimator,
\smash{$\htheta^{(OLS)} = \xi^\top (\bxcc^\top \bxcc)^{-1} \bxcc^\top Y$}, satisfies
\begin{equation}
\label{eq:ols_characterization}
\sqrt{\nc} \p{\htheta^{(OLS)} - \theta} \,\big / \, \sqrt{\xi^\top \Sigma_{\rm c}^{-1} \xi} \Rightarrow \nn\p{0, \, \sigma^2}, \ \eqand
\sqrt{\nc} \p{\htheta^{(OLS)} - \theta - \sum_{\cb{i : W_i = 0}} \gamma^*_i \varepsilon_i(0)} \rightarrow_p 0,
\end{equation}
where $\gamma^*_i$ is as defined in Lemma \ref{lemm:OLS_balance}. By comparing this characterization to our
result in Theorem \ref{theo:debiased_prediction}, it becomes apparent that our debiased estimator \smash{$\htheta$}
has been able to recover the large-sample qualitative behavior of \smash{$\htheta^{(OLS)}$}, despite being
in a high-dimensional $p \gg n$ regime.
The connection between debiasing and OLS ought not appear too surprising. After all, under classical assumptions,
\smash{$\htheta^{(OLS)}$} is known to be the minimum variance unbiased linear estimator for $\theta$; while the weights
$\gamma$ in \eqref{eq:gamma_bdd} were explicitly chosen to minimize the variance of \smash{$\htheta$} subject
to the estimator being nearly unbiased.

A downside of the above result is that our main goal is to estimate \smash{$\mucc  = \hatx \cdot \betacc$},
and this contrast-defining vector \smash{$\hatx$} fails to satisfy the bound on \smash{$\hatx^\top \Sigma^{-1} \hatx$}
assumed in Theorem \ref{theo:debiased_prediction}. In fact, because \smash{$\hatx$} is random, this quantity
will in general be on the order of $p/n$.
In the result below, we show how to get around this problem under the weaker assumption that
\smash{$m_{\rm t}^\top \Sigma_{\rm c}^{-1} m_{\rm t}$} is bounded; at a high level, the proof shows that the
the stochasticity \smash{$\hatx$} does not invalidate our previous result.
We note that, because \smash{$\bY_{\rm t}$} is uncorrelated with \smash{$\hmucc$} conditionally on \smash{$\hatx$}, the following result also
immediately implies a central limit theorem for \smash{$\htau = \bY_{\rm t} - \hmucc$}
where \smash{$\bY_{\rm t}$} is the average of the treated outcomes.

\begin{coro}
\label{coro:debiased_catt}
Under the conditions of Theorem \ref{theo:debiased_prediction}, suppose that we want to estimate
$\mucc  = \hatx \cdot \betacc$ by replacing $\xi$ with $\hatx$ in \eqref{eq:htheta}, and let
$m_{\rm t} = \EE{X \cond W = 1}$. Suppose, moreover, that we replace all the assumptions made about
$\xi$ in Theorem \ref{theo:debiased_prediction} with the following assumptions: throughout our sequence
of problems, the vector $m_{\rm t}$ satisfies $m_{\rm t} \Sigma_{\rm c}^{-1} m_{\rm t} \leq V$ and $\Norm{m_{\rm t}}_\infty \geq \kappa$.
Suppose, finally, that $(X_i - m_{\rm t})_j \cond W_i = 1$ is sub-Gaussian with parameter $\nu^2 > 0$, and that
the overall odds of receiving treatment $\PP{W = 1}/\PP{W=0}$ tend to a limit $\rho$ bounded
away from 0 and infinity. Then, setting the tuning parameter in \eqref{eq:gamma_bdd} as
$K = C\varsigma^2 \sqrt{VS} +  \nu \sqrt{2.1 \, \rho}$, we get
\begin{equation}
\label{eq:debiased_prediction_catt}
\p{\hmucc - \mucc}\,\big/\,{\Norm{\gamma}_2} \Rightarrow \nn\p{0, \, \sigma^2}, \ \  \nc \Norm{\gamma}_2^2 \,\big/\,  m_{\rm t}^\top \Sigma_{\rm c}^{-1} m_{\rm t} \leq 1 + o_p(1).
\end{equation}
\end{coro}

The asymptotic variance bound $m_{\rm t}^\top \Sigma_{\rm c}^{-1} m_{\rm t}$ is exactly the Mahalanobis distance
between the mean treated and control subjects with respect to the covariance of the control sample.
Thus, our result shows that we can achieve asymptotic inference about $\tau$ with a $1/\sqrt{n}$ rate of convergence,
irrespective of the dimension of the features, subject only to a requirement on the Mahalanobis distance between the
treated and control classes, and comparable sparsity assumptions on the $Y$-model
 as used by the rest of the high-dimensional inference literature, including
 \citet{belloni2014inference,belloni2013program}, \citet{chernozhukov2016double} and \citet{farrell2015robust}.
 However, unlike this literature, we make no assumptions on the propensity model beyond overlap, and do not
 require it to be estimated consistently. In other words, by relying more heavily on linearity of the outcome function,
 we can considerably relax the assumptions required to get $\sqrt{n}$-consistent treatment effect estimation.

\subsection{A Robust Analysis with Overlap}
\label{sec:direct}

Our discussion so far, leading up to Corollary \ref{coro:debiased_catt}, gives a characterization
of when and why we should expect approximate residual balancing to work. However, from a practical
perspective, the assumptions used in our derivation---in particular the transformed independence design
assumption---were stronger than ones we may feel comfortable making in applications.

In this section, we propose an alternative analysis of approximate residual balancing based on overlap.
Informally, overlap requires that each unit have a positive probability of receiving each of the treatment
and control conditions, and thus that the treatment and control populations cannot be too dissimilar.
Without overlap, estimation of average treatment effects relies fundamentally on extrapolation beyond
the support of the features, and thus makes estimation inherently sensitive to functional form assumptions;
and, for this reason, overlap has become a common assumption
in the literature on causal inference from observational studies \citep{crump, imbens2015causal}.
For estimation of the average effect for the treated we in fact only need the propensity score to be bounded from above by $1-\eta$,
but for estimation of the overall average effect we would require both the lower and upper bound on the propensity score.
If we are willing to assume overlap, we can relax the transposed independence design assumption into much more routine
regularity conditions on the design, as in Assumption \ref{assu:tech}.

\begin{assu}[Overlap]
\label{assu:overlap}
There is a constant $0 < \eta $ such that $\eta \leq e(x) \leq 1 - \eta$ for all $x \in \RR^p$.
\end{assu}

\begin{assu}[Design]
\label{assu:tech}
Our design $X$ satisfies the following two conditions. First, the design is sub-Gaussian, i.e.,
there is a constant $\nu > 0$ such that the distribution of $X_j$ conditional on $W = w$
is sub-Gaussian with parameter $\nu^2$ after re-centering.
Second, we assume that $\bxcc$ satisfies the $\cb{k, \, \omega, \, 4}$-restricted eigenvalue
condition as defined in Assumption \ref{assu:re}, with probability tending to 1.
\end{assu}

Following Lemma \ref{lemm:OLS_balance}, our analysis again proceeds by guessing a feasible
solution to our optimization problem, and then using it to bound the variance of our estimator.
Here, however, we use inverse-propensity weights as our guess: $\gamma^*_i \propto e(X_i)/(1 - e(X_i))$.
Our proof hinges on showing that the actual weights we get from the optimization problem are
at least as good as these inverse-propensity weights, and thus our method will be at most as variable
as one that uses augmented inverse-propensity weighting \eqref{eq:aipw} with these oracle propensity weights.

\begin{theo}
\label{theo:overlap}
Suppose that we have $n$ independent and identically distributed training examples satisfying
Assumptions \ref{assu:unconf}, \ref{assu:linearity}, \ref{assu:spar}, \ref{assu:overlap}, \ref{assu:tech},
and that the treatment odds $\PP{W = 1}/\PP{W=0}$ converge to $\rho$ with $0 < \rho < \infty$.
Suppose, moreover, that we have homoskedastic noise:
\smash{$\operatorname{Var}[\varepsilon_i(w) \cond X_i] = \sigma^2$} for all $i = 1, \, ..., \, n$,
\sloppy{and also that the response noise \smash{$\varepsilon_i(w) := Y_i(w) - \mathbb{E}[Y_i(w) \cond X_i]$} is
uniformly sub-Gaussian with parameter $\upsilon^2 > 0$.}
Finally, suppose that we use \eqref{eq:meta_alg} with weights \eqref{eq:gamma_def},
except we replace the Lagrange-form penalty on the imbalance with a hard constraint
\smash{$\lVert \hatx - \bxcc^\top \tgamma\rVert_\infty \leq K \sqrt{\log(p)/n_{\rm c}}$},
with \smash{$K = \nu\sqrt{2.1(\rho + (\eta^{-1}-1)^2}$}.
Moreover, we fit the outcome model using a lasso with penalty parameter
set to \smash{$\lambda_n = 5 \nu \upsilon \sqrt{\log\p{p} / \nc}$}. Then,
\begin{align}
\label{eq:gauss1}
\frac{\hmucc - \mucc}{\Norm{\gamma}_2}  \ \Rightarrow \ \nn\p{0, \, \sigma^2} \ \eqand \  \frac{\htau - \tau}{\sqrt{\nt^{-1} + \Norm{\gamma}_2^2}}  \ \Rightarrow \ \nn\p{0, \, \sigma^2},
\end{align}
where $\tau$ is the expected treatment effect on the treated  \eqref{eq:taus}.
Moreover,
\begin{equation}
\label{eq:inference_rate}
 \limsup_{n \rightarrow \infty} \ \nc \, \Norm{\gamma}_2^2 \ \leq \ \rho^{-2} \, \EE{\left.\p{\frac{e\p{X_i}}{1 - e\p{X_i}}}^2 \right| W_i = 0}.
\end{equation}
\end{theo}

The rate of convergence guaranteed by \eqref{eq:inference_rate} is the same as what we would get if we
actually knew the true propensities and could use them for weighting \citep{robins1994estimation,robins2}. Here, we
achieve this rate although we have no guarantees that the true propensities $e(X_i)$ are consistently estimable.
Finally, we note that, when the assumptions to Corollary \ref{coro:debiased_catt} hold, the bound
\eqref{eq:debiased_prediction_catt} is stronger than \eqref{eq:inference_rate}; however, there exist
designs where the bounds match \citep{wang2017approximate}.

Finally, in applications, it is often of interest to have confidence intervals for $\mucc$ and $\tau$ rather than just point estimates;
below, we propose such a construction. Much like the sandwich variance estimates for ordinary least squares regression,
our proposed confidence intervals are heteroskedasticity robust even though the underlying point estimates were motivated
using an argument written in terms of a homoskedastic sampling distribution.

\begin{coro}
\label{coro:inference}
Under the conditions of Theorems \ref{theo:debiased_prediction} or \ref{theo:overlap}, suppose instead that we have heteroskedastic noise
$\upsilon_{\min}^2 \leq \Var{\varepsilon_i(W_i) \cond X_i, \, W_i} \leq \upsilon^2 \ \text{ for all } \ i = 1, \, ..., \, n$.
Then, the following holds:
\begin{equation}
\label{eq:pivot}
\p{\hmucc - \mucc} \,\Big/\, \sqrt{\hV_{\rm c}} \Rightarrow \nn\p{0, \, 1}, \ \ \hV_{\rm c} = \sum_{\cb{i : W_i = 0}} \gamma_i^2 \p{Y_i - X_i \cdot \hbetacc}^2.
\end{equation}
\end{coro}

In order to provide inference about $\tau$, we also need error bounds for $\hmutt$. 
Under sparsity assumptions comparable to those made for $\betacc$ in Theorem \ref{theo:overlap}, we can verify that
\begin{equation}
 (\hmutt - \mutt) \,\Big/\, \sqrt{\hV_{\rm t} } \Rightarrow \p{0, \, 1}, \ \ \hV_{\rm t} = \frac{1}{\nt^2} \sum_{\cb{i : W_i = 1}} \p{Y_i - X_i \hbetatt}^2,
\end{equation}
where $\hbetatt$ is obtained using the lasso with $\lambda_n = 5 \nu\upsilon \sqrt{\log\p{p} / \nc}$. Moreover, $\hmucc$ and $\hmutt$ are independent conditionally on $X$ and $W$, thus implying that \smash{$ (\htau - \tau) \, /\, (\hV_{\rm c} + \hV_{\rm t})^{1/2} \Rightarrow \nn\p{0, \, 1}$}. This last expression is what we use for building confidence intervals for $\tau$.

\section{Application: The Efficacy of Welfare-to-Work Programs}

Starting in 1986, California implemented the Greater Avenues to Independence (GAIN) program,
with an aim to reduce dependence on welfare and promote work among disadvantaged households.
The GAIN program provided its participants with a mix of educational resources such as
English as a second language courses and vocational training, and job search assistance.
This program is described in detail by \citet{hotz2006evaluating}.
In order to evaluate the effect of GAIN, the Manpower Development Research Corporation conducted
a randomized study between 1988 and 1993, where a random subset of GAIN registrants were eligible
to receive GAIN benefits immediately, whereas others were embargoed from the program until 1993
(after which point they were allowed to participate in the program). All experimental subjects were
followed for a 3-year post-randomization period.

The randomization for the GAIN evaluation was conducted separately by county; following
\citet{hotz2006evaluating}, we consider data from Alameda, Los Angeles, Riverside and
San Diego counties. As discussed in detail in \citet{hotz2006evaluating}, the experimental
conditions differed noticeably across counties, both in terms of the fraction of registrants
eligible for GAIN, i.e., the treatment propensity, and in terms of the subjects participating in
the experiment. For example, the GAIN programs in Riverside and San Diego counties sought
to register all welfare cases in GAIN, while the programs in Alameda and Los Angeles counties
focused on long-term welfare recipients.

\begin{figure}[t]
\centering
\begin{tabular}{cc}
\includegraphics[width=0.4\textwidth, trim=2mm 8mm 2mm 10mm, clip=TRUE]{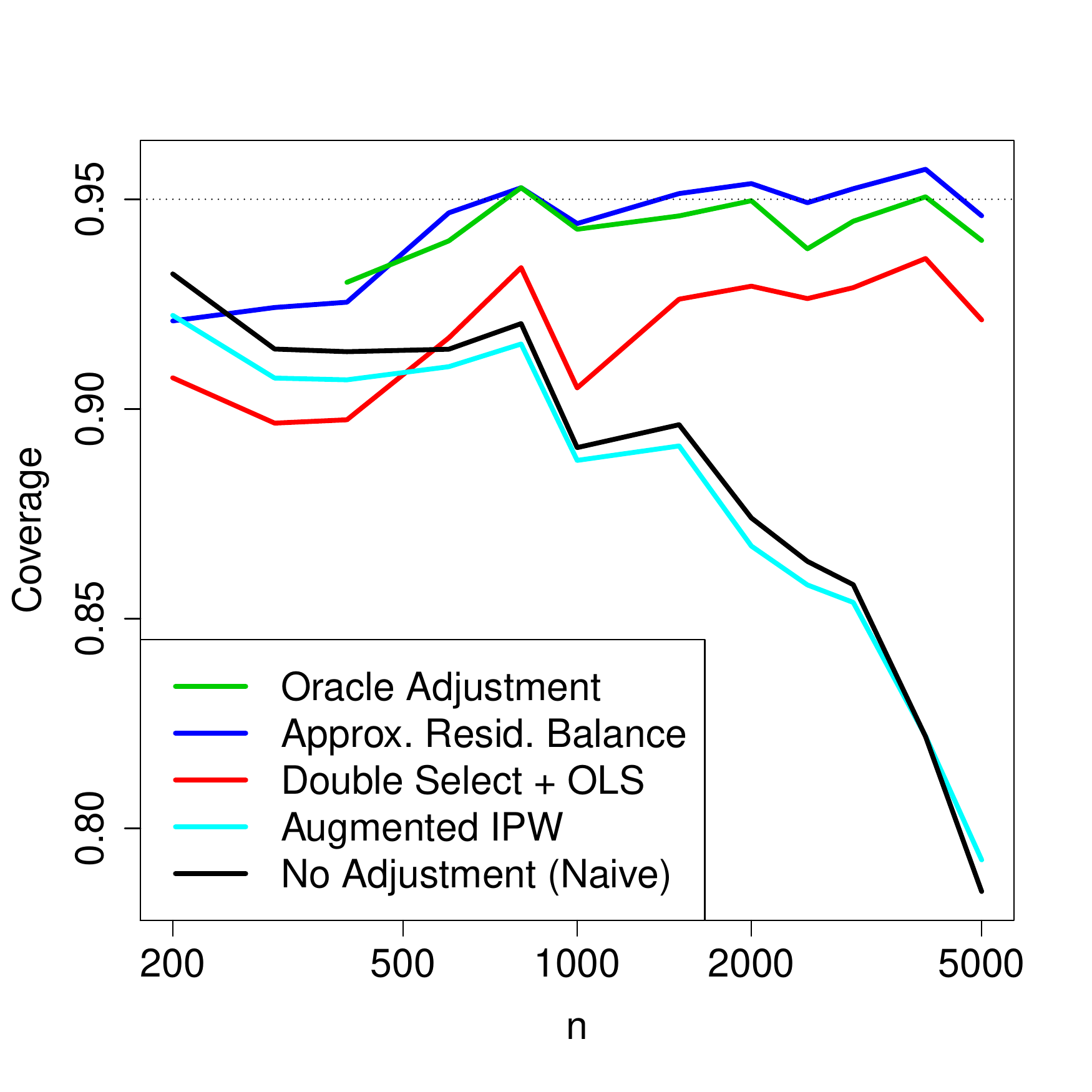} &
\includegraphics[width=0.4\textwidth, trim=2mm 8mm 2mm 10mm, clip=TRUE]{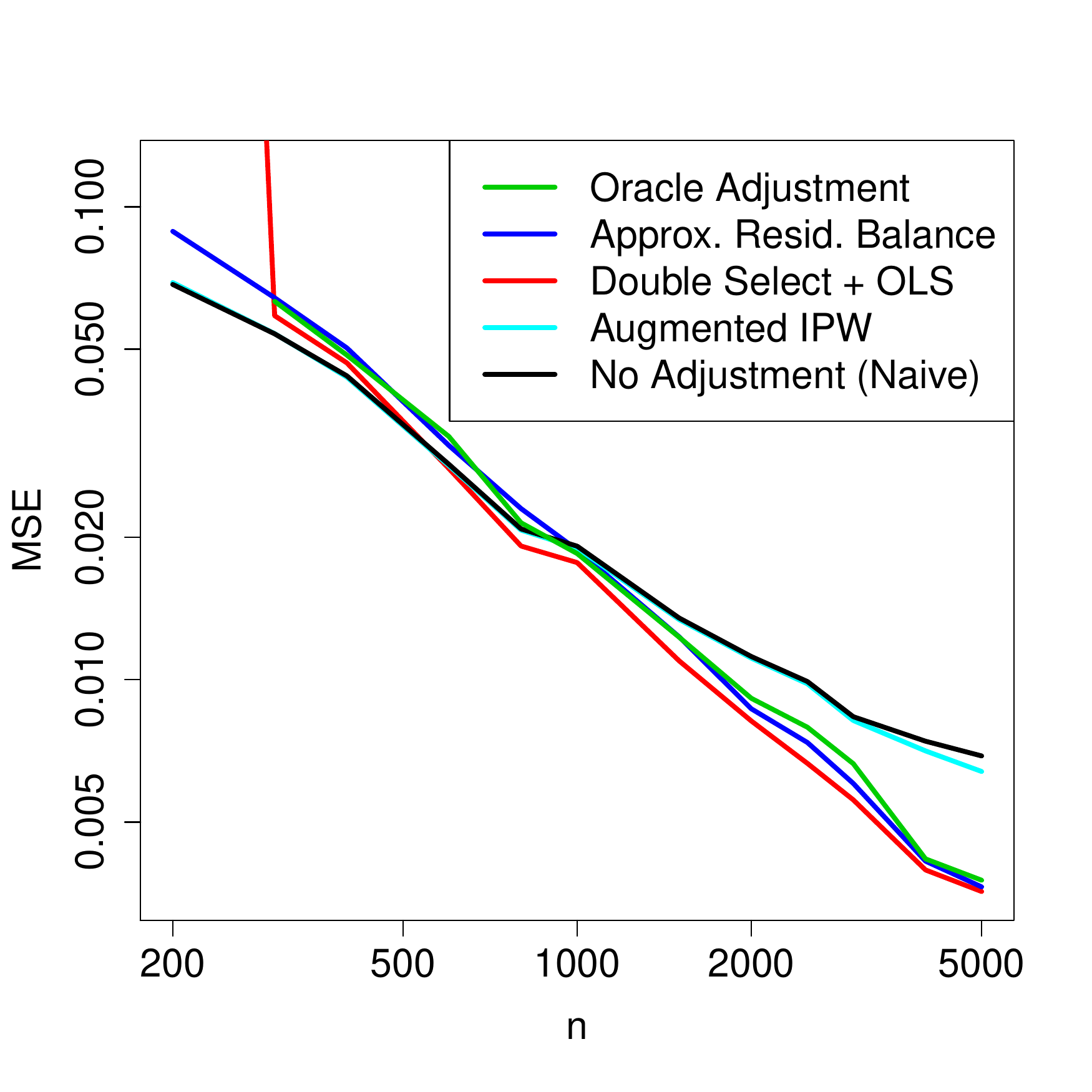} \\
\text{Coverage} & \text{Mean-Squared Error}
\end{tabular}
\caption{Finite sample performance of the average treatment effect on the treated for different estimators,
aggregated over 1,000 replications. The target coverage rate, 0.95, is denoted with a dotted line.}
\label{fig:labor}
\end{figure}

The fact that the randomization of the GAIN evaluation was done at the county level rather than
at the state level presents us with a natural opportunity to test our method, as follows. We seek
to estimate the average treatment effect of GAIN on the treated; however, we hide the county
information from our procedure, and instead try to compensate for sampling bias by controlling
for a large amount of covariates. We used spline expansions of age and prior income, indicators
for race, family status, etc., for a total of $p = 93$ covariates. Meanwhile, we can check our
performance against a gold standard estimate of the average treatment effect that is stratified by
county and thus guaranteed to be unbiased.\footnote{More formally, in our experiments, we
set the gold standard using the county-stratified oracle estimator on bootstrap samples
of the full $n = 19,170$ sample. We use bootstrap samples to correct for the correlation of estimators
$\htau$ obtained using the full dataset and subsamples of it. We also note that, given this setup,
the quantity we are using as our gold standard is not and estimate of $\tau$, i.e., the conditional average
treatment effect on the treated sample, and should rather be thought of as an estimate of $\EE{\tau}$, i.e.,
the average treatment effect on the treated population. Since we are in a setting with a fairly weak signal,
this should not make a noticeable difference in practice.}

We compare the behavior of different methods for estimating the average treatment
effect on the treated using randomly drawn subsamples of the original data (the full dataset has $n = 19,170$).
In addition to approximate residual balancing, we consider augmented inverse-propensity
weighting \eqref{eq:aipw} and double selection following \citet{belloni2014inference} as our baselines.
We also show the behavior of an ``oracle'' procedure
that gets to observe the hidden county information and then simply estimates treatment effects for each
county separately, and the ``naive'' difference-in-means estimator that ignores the features $X$. In very small samples,
the oracle procedure is not always well defined because some samples may result in counties where either
everyone or no one is treated.

Figure \ref{fig:labor} compares the performance of the different methods.
We see that approximate residual balancing and double selection both do well in terms of mean-squared error.
Moreover, confidence intervals built via approximate residual balancing achieve effectively nominal coverage;
double selection also gets reasonable coverage and improves with $n$.
In contrast, augmented inverse-propensity weighting does not perform well here. The problem appears to be that
estimating treatment propensities is quite difficult, and a cross-validated logistic elastic net often
learns an effectively constant propensity model.

\section{Simulation Experiments}
\label{sec:simu}

\subsection{Methods under Comparison}

In addition to {\bf approximate residual balancing} as described in Procedure \ref{alg:approx_balance}, the methods we use as baselines are as follows:
{\bf naive} difference-in-means estimation $\htau = \overline{Y}_{\rm t} - \overline{Y}_{\rm c}$ that ignores the covariate information $X$;
the {\bf elastic net} \citep{zou2005regularization}, or equivalently, Procedure \ref{alg:approx_balance} with trivial weights $\gamma_i = 1/\nc$;
{\bf approximate balancing}, or equivalently, Procedure \ref{alg:approx_balance} with trivial parameter estimates \smash{$\hbetacc = 0$} \citep{zubizarreta2015stable};
{\bf inverse-propensity weighting}, as discussed in Section \ref{sec:weighted_avg}, with propensity estimates $\he(X_i)$ obtained by elastic net logistic regression, with the propensity scores trimmed at $0.05$ and $0.95$;
{\bf augmented inverse-propensity weighting}, which pairs elastic net regression adjustments
with the above inverse-propensity weights \eqref{eq:aipw};
the {\bf weighted elastic net}, motivated by \citet{schafer}, that uses inverse-propensity weights as
sample weights for the elastic net regression;
{\bf targeted maximum likelihood estimation} (TMLE), which fine-tunes the elastic net
regression estimates along the direction specified by the inverse-propensity weights \citep{van2006targeted},
and
{\bf ordinary least squares after model selection} where, in the spirit of \citet{belloni2014inference}, we run lasso linear regression for $Y \cond X, \, W = 0$
and lasso logistic regression for $W \cond X$, and then compute the ordinary least squares estimate for $\tau$ on the union of the support of the three lasso problems.

Unless otherwise specified, all outcome and propensity models were fit using a (linear or logistic) elastic
net.
Whenever there is a ``$\lambda$'' regularization parameter to be selected, we use cross validation with the \texttt{lambda.1se} rule from the \texttt{glmnet} package \citep{friedman2010regularization}. In \citet{belloni2014inference}, the authors recommend selecting $\lambda$ using more sophisticated methods, such as the square-root lasso \citep{belloni2011square}. However, in our simulations, our implementation of \citet{belloni2014inference} still attains excellent performance in the regimes the method is designed to work in.
Similarly, our confidence intervals for $\tau$ are built using a cross-validated choice of $\lambda$ instead of the fixed choice assumed by Corollary \ref{coro:inference}.
Our implementation of approximate residual balancing, as well as all the discussed baselines, is available in the \texttt{R}-package \texttt{balanceHD}.

\subsection{Simulation Designs}

\begin{figure}[t]
\centering
  \begin{subfigure}[b]{0.4\textwidth}
            \centering
            \includegraphics[width=\textwidth]{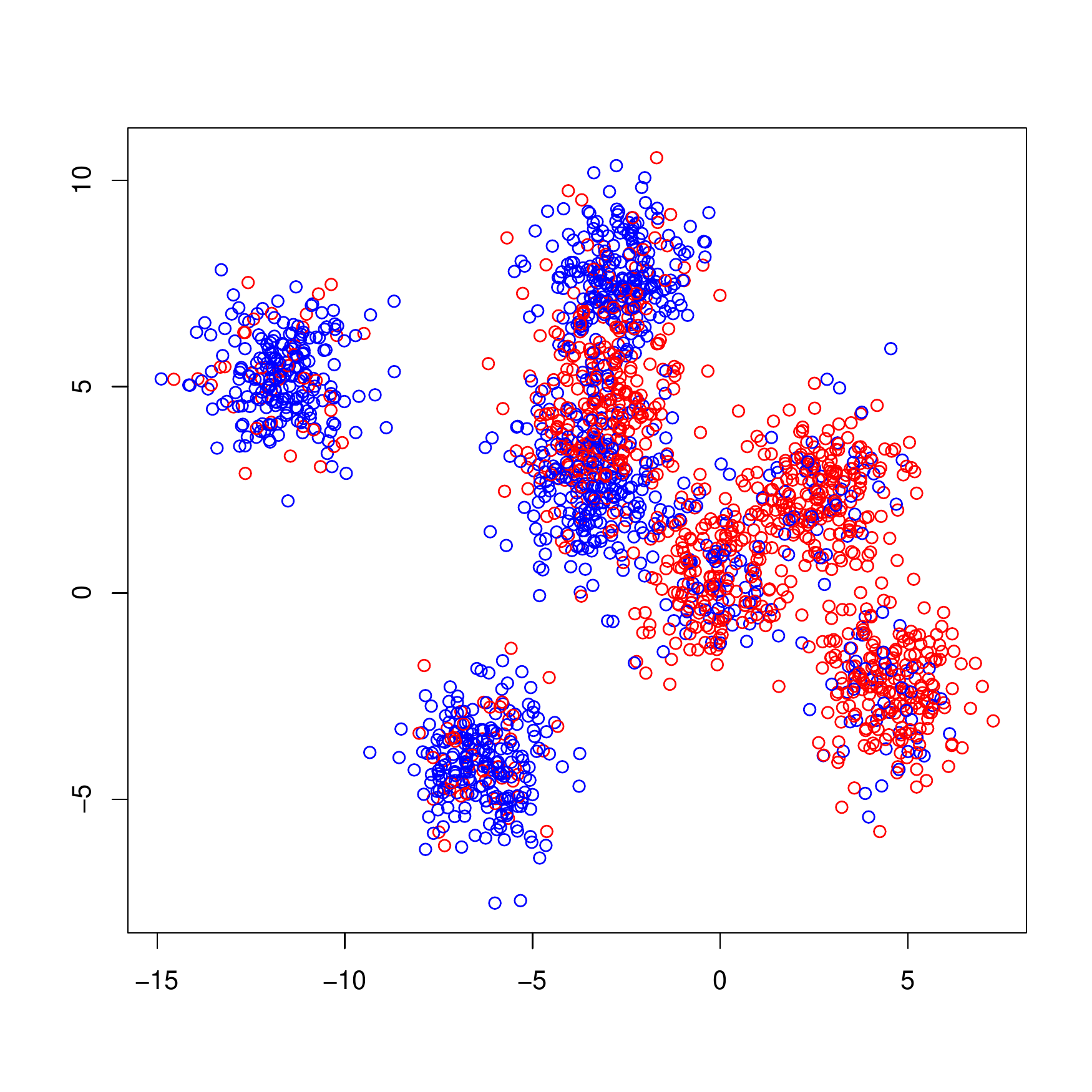}
            \caption{Low-dimensional version of the many clusters simulation setting. The blue and red dots denote control and treated $X$-observations respectively.}
    \label{fig:manyclust}
    \end{subfigure}
    \hspace{0.05\textwidth}
    \begin{subfigure}[b]{0.4\textwidth}
            \centering
            \includegraphics[width=\textwidth]{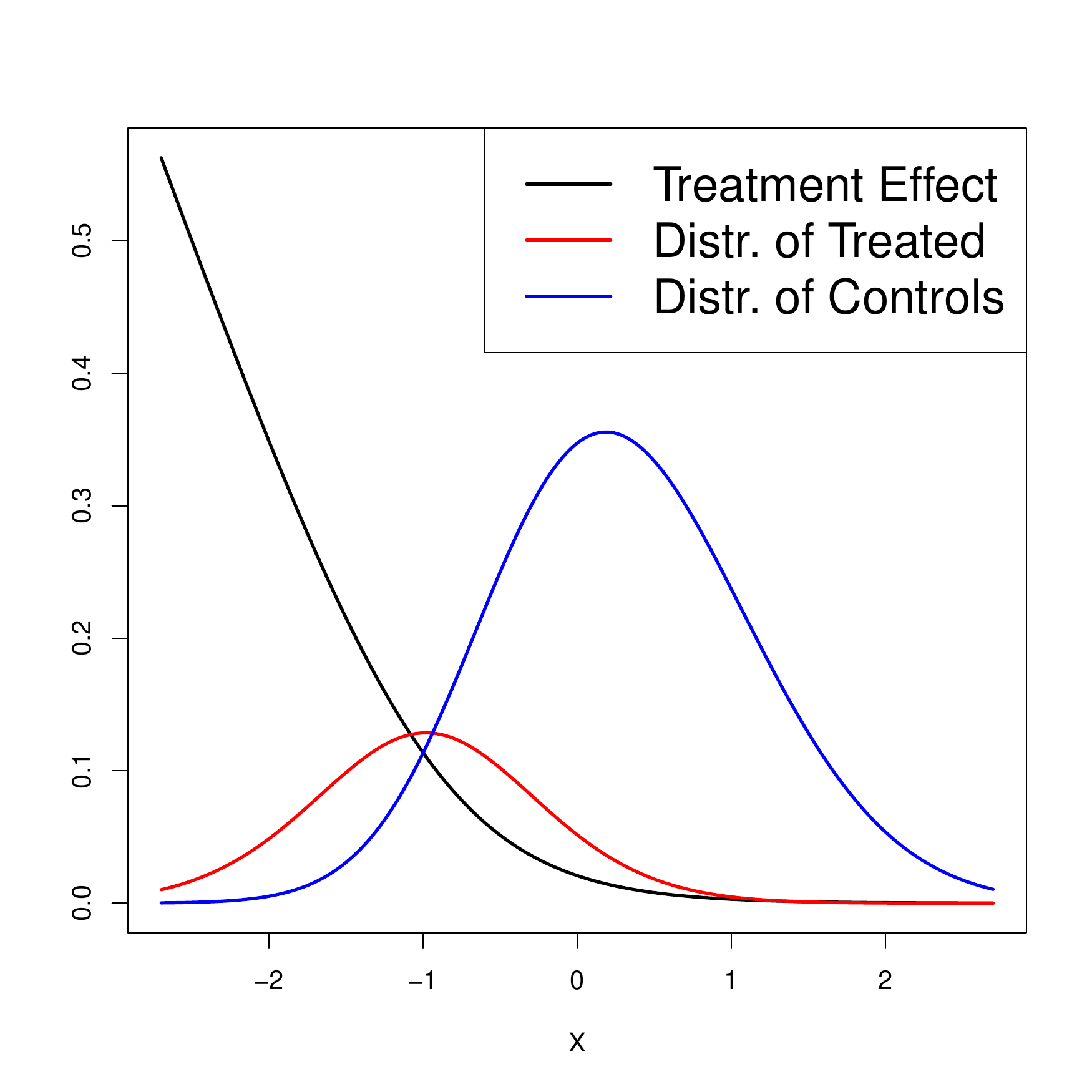}
            \caption{Schematic of misspecified simulation setting, along the first covariate $(X_i)_1$. The ``treatment effect'' curve is not to scale along the $Y$-axis.}
    \label{fig:misspec}
    \end{subfigure}
    \caption{Illustrating simulation designs.}
\end{figure}

We consider five different simulation settings. Our first setting is a {\bf two-cluster} layout, with data drawn
as $Y_i = (C_i + Z_i) \cdot \beta + W_i + \varepsilon_i$. Here,
$W_i = \text{Bernoulli}(0.5)$,
$Z_i \sim \nn\p{0, \, I_{p \times p}}$,
$\varepsilon_i \sim \nn\p{0, \, 1}$,
and $C_i \in \RR^p$ is a cluster center that is one of $C_i \in \cb{0, \, \delta}$,
such that $\PP{C_i = 0 \cond W_i = 0} = 0.8$ and $\PP{C_i = 0 \cond W_i = 1} = 0.2$.
We consider two settings for the between-cluster vector $\delta$: a ``dense'' setting where
$\delta = 4/\sqrt{n} \ \mathbf{1}$, and a ``sparse'' setting where
$\delta_j = 40/\sqrt{n} \ 1\p{\cb{j = 1 \text{ modulo } 10}}$.
Our second {\bf many-cluster} layout is closely related to the first, except now we have 20 cluster
centers $C_i \in \cb{c_1, \, ..., \, c_{20}}$, where all the cluster centers are independently
generated as $c_k \sim \nn\p{0, \, I_{p \times p}}$. To generate the data, we first draw
$C_i$ uniformly at random from one of the 20 cluster centers and then set $W_i = 1$ wit
probability $\eta$ for the first 10 clusters and $W_i = 1$ with probability $1 - \eta$ for the last 10 clusters;
we tried both $\eta = 0.1$ and $\eta = 0.25$.
We illustrate this simulation concept in Figure \ref{fig:manyclust}.
In both cases, we chose $\beta$ as one of
\begin{equation}
\label{eq:betas}
\begin{split}
&\text{dense}: \beta \propto (1, \, {1}/{\sqrt{2}}, \, ..., \, {1}/{\sqrt{p}}), \ \
\text{harmonic}: \beta \propto \p{{1}/{10}, {1}/{11}, ..., {1}/\p{p+9}}, \\
&\text{moderately sparse}: \beta \propto (\underbrace{10, \, ..., \, 10}_{10}, \, \underbrace{1, \, ..., \, 1}_{90}, \, \underbrace{0, \, ..., \, 0}_{p - 100}),  \eqand 
\text{very sparse}: \beta \propto (\underbrace{1, \, ..., \, 1}_{10}, \,  \underbrace{0, \, ..., \, 0}_{p - 10}).
\end{split}
\end{equation}
The signal strength was scaled such that $\Norm{\beta}_2 = 2$ in the two-cluster layout
and $\Norm{\beta}_2 = 3$ in the many-cluster layout.

Our next two simulations are built using more traditional structural models.
We first consider a {\bf sparse two-stage} setting closely inspired by an experiment of
\citet{belloni2014inference}. Here $X_i \sim \nn\p{0, \, \Sigma}$ with $\Sigma_{ij} = \rho^{\abs{i - j}}$, and
$\theta_i = X_i \cdot \beta_W + \varepsilon_{i1}$. Then, $W_i \sim \text{Bernoulli}(1/(1 + e^{\theta_i}))$, and
finally $Y_i = X_i \cdot \beta_Y + 0.5 \, W_i + \varepsilon_{i2}$ where $\varepsilon_{i1}$ and $\varepsilon_{i2}$
are independent standard Gaussian. Following \citet{belloni2014inference}, we set the structure model as
$(\beta_Y)_j \propto 1/j^2$ for $j = 1, \, ..., \, p$; for the propensity model, we consider both a ``very sparse''
propensity model $(\beta_W)_j \propto 1/j^2$, and also a ``dense'' propensity model $(\beta_W)_j \propto 1/\sqrt{j}$. 
A potential criticism of this simulation design is that the signal is perhaps unusually sparse
(in the 4-th column of Table \ref{tab:bch}, adjusting for differences in the two most important covariates
removes 93\% of the bias associated with all the covariates); moreover, we note that all the important
coefficients of both $\beta_Y$ and $\beta_W$ are close to each other in terms of their indices; thus, the
effect of using a correlated design may be mitigated.
Thus, we also ran a {\bf moderately sparse two-stage} simulation, just like the above one, except we now used
choices for $\beta_Y$ as in \eqref{eq:betas}, the only difference being that we shifted the indices of the betas,
multiplying them by 23 mod $p$ (e.g., the harmonic setup now has $(\beta)_j \propto 1/[10 + (23\,(j - 1) \text{ mod } p)]$).
Here, we drew the treatment assignments from a well-specified logistic model,
$W_i \sim \text{Bernoulli}(1/(1 + \exp(-\sum_{j = 1}^{100} X_{ij} / 40)))$.

To test the robustness of all considered methods, we also ran a {\bf misspecified} simulation. Here, we first drew $X_i \sim \nn\p{0, \, I_{p \times p}}$, and defined latent parameters $\theta_i = \log(1 + \exp(-2 - 2 * (X_i)_1)) / 0.915$. We then drew $W_i \sim \text{Bernoulli}(1 - e^{-\theta_i})$, and finally $Y_i = (X_i)_1 + \cdots + (X_i)_{10} + \theta_i  (2W_i - 1)/2 + \varepsilon_i$ with $\varepsilon_i \sim \nn\p{0, \, 1}$. We varied $n$ and $p$.
This simulation setting, loosely inspired by the classic program evaluation dataset of \citet{lalonde1986evaluating}, is illustrated in Figure \ref{fig:misspec}; note that the average treatment effect on the treated is much greater than the overall average treatment effect here.

\subsection{Results}

In the first two experiments, for which we report results in Tables \ref{tab:first} and \ref{tab:manyclust},
the outcome model $Y|X$ is reasonably sparse, while the propensity model has overlap but is not in
general sparse. In relative terms, this appears to hurt the double-selection method most. 
Meanwhile, in Table \ref{tab:bch}, we find that the method of \citet{belloni2014inference} has excellent
performance---as expected---when both the propensity and outcome models are sparse. However,
if we make the problem somewhat more difficult (Table \ref{tab:ar}), its performance decays substantially,
and double selection lags both approximate residual balancing and propensity-based methods in its performance.

Generally, we find that the balancing performs substantially better than propensity score weighting,
with or without direct covariate adjustment. We also find that combining direct covariate adjustment
with weighting does better than weighting on its own, irrespective of whether the weighting is based
on balance or on the propensity score. In these experiments, the weighted elastic net and TMLE also somewhat
improve over AIPW.

Encouragingly, approximate residual balancing also does a good job in the misspecified setting from Table \ref{tab:misspec}.
It appears that our stipulation that the approximately balancing weights \eqref{eq:gamma_def} must be non-negative
(i.e., $\gamma_i \geq 0$) helps prevent our method from extrapolating too aggressively. Conversely, least squares
with model selection does not perform well despite both the outcome and propensity models being sparse; apparently,
it is more sensitive to the misspecification here. Perhaps the reason AIPW and TMLE do not do as well here is that there
are very strong linear effects.

We evaluate coverage of confidence intervals in the ``many-cluster'' setting for different choices of $\beta$, $n$, and $p$;
results are given in Table \ref{tab:coverage}. Coverage is generally better with more overlap ($\eta = 0.25$) rather than
less ($\eta = 0.1$), and with sparser choices of $\beta$. Moreover, coverage rates appear to improve as $n$ increases,
suggesting that we are in a regime where the asymptotics from Corollary \ref{coro:inference} are beginning to apply.

\section{Discussion}

In this paper, we introduced approximate residual balancing as a method for unconfounded average treatment effect estimation
in high-dimensional linear models. Under standard assumptions from the high-dimensional
inference literature, our method allows for $\sqrt{n}$-consistent inference of the average treatment effect without
any structural assumptions on the treatment assignment mechanism beyond overlap.

Widely used doubly robust methods, pioneered by \citet{robins1994estimation} and studied further by several
authors \citep[e.g.,][]{belloni2013program,farrell2015robust,schafer,
scharfstein1999adjusting,robins2007comment,tan2010bounded,van2006targeted}, approach this problem
by trying to estimate two different nuisance components, the outcome model and the propensity model.
These methods then achieve consistency if either nuisance component is itself consistently estimated, and
achieve semiparametric efficiency if both components as estimated fast enough.
In contrast, our method ``bets'' on linearity twice, both in fitting the lasso and in attempting to
balance away its bias. In well specified linear models, this bet allows us to considerably extend the
class of problems for with $\sqrt{n}$-consistent inference of average treatment effects is possible;
thus, if a practitioner believes linearity to be a reasonable assumption in a given problem,
our estimator may be a promising choice.

We end by mentioning two important questions left open by this paper. First, it would be important
to develop a better understanding of how to choose the tuning parameter $\zeta$ in
\eqref{eq:gamma_def} that trades off bias and variance in our balancing. Results from Theorems \ref{theo:debiased_prediction}
and \ref{theo:overlap} provide some guidance on choosing $\zeta$ (via a constraint-form characterization);
however, in our experiments, we achieved good performance by simply setting $\zeta = 1/2$ everywhere.
The difficulty in choosing $\zeta$ is that we are trying to trade off an observable quantity (sampling variance)
against an unobservable one (residual bias), and so cannot rely on simple methods like cross-validation
that require unbiased estimates of the loss criterion we are trying to minimize. It would be of considerable interest
to either devise a data-adaptive choice for $\zeta$, or understand why a fixed choice $\zeta = 1/2$ appears
to achieve systematically good performance.

\sloppy{
It would also be interesting to extend our approach to generalized linear models, where
there is a non-linear link function $\psi$ for which \smash{$\mathbb{E}[Y_i(c) \cond X_i = x] = \psi(x \cdot \betacc)$}.
In causal inference applications, this setting frequently arises when the outcomes $Y_i^\obs$ are binary,
and we are willing to work with a logistic regression model. In this case, the first-order error
component from using a pilot estimator \smash{$\hbetacc$}
for estimating $\mucc$ with a plug-in estimator
\smash{$ n_{\rm c}^{-1}\sum_{\{W_i = 1\}}\psi(X_i \cdot \hbetacc)$}
would be of the form
\smash{$n_{\rm c}^{-1} \sum_{\{W_i = 1\}}\psi'(X_i \cdot \hbetacc)X_i(\betacc - \hbetacc)$}.
An analogue to Proposition \ref{prop:main} then suggests using an estimator}
\begin{equation}
\begin{split}
&\hmucc = \frac{1}{n_{\rm c}} \sum_{\{W_i = 1\}}\psi\p{X_i \cdot \hbetacc} + \sum_{\{W_i = 0\}} \gamma_i \p{Y_i^\obs - \psi\p{X_i \cdot \hbetacc}}, \\
&\gamma = \argmin_{\tgamma}\Bigg\{ \zeta \Norm{\frac{1}{n_{ \rm t}} \sum_{\{W_i = 1\}}\psi'\p{X_i \cdot \hbetacc}X_i -  \sum_{\{W_i = 0\}} \tgamma_i \, \psi'\p{X_i \cdot \hbetacc}X_i}_\infty^2  \\
&\ \ \ \ \ \ \ \ \ \ \ \ \ \ +  \p{1 - \zeta} \sum_{\cb{W_i = 0}} \tgamma_i^2 \, \psi'\p{X_i \cdot \hbetacc} 
\text{ subject to } \sum_{\cb{W_i = 0}} \tgamma_i = 1, \ \ 0 \leq \tgamma_i \leq n_{\rm c}^{-2/3} \Bigg\}.
\end{split}
\end{equation}
However, due to space constraints, we leave a study of this estimator to further work.

{\small
\setlength{\bibsep}{0.2pt plus 0.3ex}
\bibliographystyle{plainnat-abbrev}
\bibliography{references}
}


\newpage

\begin{table}[p]
\centering
\begin{tabular}{|r|cc|cc|cc|cc|}
   \hline
Beta Model & \multicolumn{2}{c|}{dense} &   \multicolumn{2}{c|}{harmonic}  & \multicolumn{2}{c|}{moderately sparse}  & \multicolumn{2}{c|}{very sparse}   \\ 
  Propensity Model & dense & sparse & dense & sparse & dense & sparse & dense & sparse \\ 
   \hline
Naive & 6.625 & 7.119 & 3.557 & 3.924 & 1.257 & 1.256 & 0.711 & 0.722 \\
  Elastic Net & 4.328 & 1.058 & 2.190 & 0.665 & 0.716 & 0.350 & 0.237 & 0.204 \\
   \hline
Approximate Balance &  \bf 3.960 & 1.179 & 2.130 & 0.686 & 0.789 & 0.362 & 0.464 & 0.316 \\
  Approx.~Residual Balance &  \bf 3.832 &  \bf 0.423 &  \bf 1.854 &  \bf 0.320 &  \bf 0.495 &  \bf 0.213 &  \bf 0.185 &  \bf 0.165 \\
   \hline
Inv.~Propensity Weight & 5.341 & 3.094 & 2.866 & 1.707 & 1.026 & 0.596 & 0.586 & 0.398 \\
  Augmented IPW & 4.082 & 0.618 & 2.031 & 0.415 & 0.607 & 0.242 & 0.209 &  \bf 0.166 \\
   \hline
Weighted Elastic Net & 4.086 & 0.562 & 1.984 & 0.385 & 0.575 & 0.232 & 0.207 &  \bf 0.171 \\
  TMLE Elastic Net &  \bf 3.811 & 0.591 &  \bf 1.843 & 0.399 &  \bf 0.495 & 0.239 &  \bf 0.192 &  \bf 0.165 \\
   \hline
Double-Select + OLS & 6.625 & 0.620 & 3.540 & 0.430 & 0.525 & 0.233 & 0.254 &  \bf 0.165 \\
   \hline
\end{tabular}

\caption{Root-mean-squared error $\sqrt{\EE{(\htau - \tau)^2}}$ in the two-cluster setting.
We used $n = 500$, $p = 2000$, and scaled the signal such that $\Norm{\beta}_2 = 2$.
All numbers are averaged over 400 simulation replications.}
\label{tab:first}
\end{table}

\begin{table}[p]
\centering

\begin{tabular}{|r|cc|cc|cc|cc|}
   \hline
Beta Model & \multicolumn{2}{c|}{dense} &   \multicolumn{2}{c|}{harmonic}  & \multicolumn{2}{c|}{moderately sparse}  & \multicolumn{2}{c|}{very sparse}   \\ 
  Overlap ($\eta$) & 0.1 & 0.25 & 0.1 & 0.25 & 0.1 & 0.25 & 0.1 & 0.25 \\ 
   \hline
Naive & 4.921 & 3.283 & 5.100 & 3.231 & 4.776 & 3.270 & 5.078 & 3.348 \\
  Elastic Net & 2.527 & 1.353 & 1.618 &  \bf 0.869 & 0.727 &  \bf 0.385 & 0.168 & 0.108 \\
   \hline
Approximate Balance & 2.575 & 1.324 & 2.505 & 1.379 & 2.567 & 1.248 & 2.434 & 1.396 \\
  Approx.~Residual Balance &  \bf 2.123 &  \bf 1.172 &  \bf 1.528 &  \bf 0.866 &  \bf 0.653 &  \bf 0.383 &  \bf 0.158 &  \bf 0.105 \\
   \hline
Inv.~Propensity Weight & 2.626 & 1.983 & 2.625 & 1.943 & 2.586 & 1.896 & 2.568 & 2.000 \\
  Augmented IPW &  \bf 2.102 & 1.233 &  \bf 1.515 &  \bf 0.852 &  \bf 0.656 &  \bf 0.376 &  \bf 0.154 &  \bf 0.103 \\
   \hline
Weighted Elastic Net &  \bf 2.061 &  \bf 1.176 & 1.576 &  \bf 0.862 & 0.727 &  \bf 0.388 & 0.194 &  \bf 0.108 \\
  TMLE Elastic Net &  \bf 2.095 &  \bf 1.208 &  \bf 1.500 &  \bf 0.847 &  \bf 0.656 &  \bf 0.375 & 0.163 &  \bf 0.103 \\
   \hline
Double-Select + OLS & 2.726 & 1.526 & 3.364 & 1.840 & 2.201 & 1.092 & 0.211 & 0.114 \\
   \hline
\end{tabular}

\caption{Root-mean-squared error $\sqrt{\EE{(\htau - \tau)^2}}$ in the many-cluster setting.
We used $n = 800$, $p = 4000$, and scaled the signal such that $\Norm{\beta}_2 = 3$.
All numbers are averaged over 400 simulation replications.}
\label{tab:manyclust}
\end{table}

\begin{table}[p]
\centering

\begin{tabular}{|r|cc|cc|cc|cc|}
   \hline
Propensity Model & \multicolumn{4}{c|}{sparse}  & \multicolumn{4}{c|}{dense}  \\ \hline
  First Stage Sig. Strength & \multicolumn{2}{c|}{$\Norm{\beta_W}_2 = 1$} & \multicolumn{2}{c|}{$\Norm{\beta_W}_2 = 4$} & \multicolumn{2}{c|}{$\Norm{\beta_W}_2 = 1$} & \multicolumn{2}{c|}{$\Norm{\beta_W}_2 = 4$}  \\ 
  Structure Sig. Strength ($\Norm{\beta_Y}_2$) & $1$ & $4$ & $1$ & $4$ & $1$ & $4$ & $1$ & $4$ \\ 
   \hline
Naive & 0.963 & 3.796 & 1.701 & 6.804 & 0.535 & 2.129 & 0.784 & 3.130 \\
  Elastic Net & 0.277 & 0.246 & 0.648 & 0.619 & 0.202 & 0.279 & 0.307 & 0.433 \\
   \hline
Approximate Balance & 0.195 & 0.662 & 0.585 & 2.313 & 0.198 & 0.731 & 0.260 & 0.987 \\
  Approx.~Residual Balance & 0.109 & 0.102 & 0.287 & 0.326 & 0.107 & 0.138 &  \bf 0.134 &  \bf 0.192 \\
   \hline
Inv.~Propensity Weight & 0.484 & 1.876 & 0.932 & 3.722 & 0.301 & 1.191 & 0.421 & 1.651 \\
  Augmented IPW & 0.164 & 0.151 & 0.374 & 0.384 & 0.130 & 0.188 & 0.181 & 0.258 \\
   \hline
Weighted Elastic Net & 0.174 & 0.163 & 0.686 & 0.708 & 0.132 & 0.193 & 0.201 & 0.300 \\
  TMLE Elastic Net & 0.161 & 0.149 & 0.234 & 0.227 & 0.122 & 0.175 & 0.355 & 0.389 \\
   \hline
Double-Select + OLS &  \bf 0.081 &  \bf 0.077 &  \bf 0.115 &  \bf 0.123 &  \bf 0.092 &  \bf 0.093 & 0.190 &  \bf 0.194 \\
   \hline
\end{tabular}

\caption{Root-mean-squared error $\sqrt{\EE{(\htau - \tau)^2}}$ in the sparse two-stage setting.
We used $n = 1000$, $p = 2000$, and $\rho = 0.5$.
All numbers are averaged over 400 simulation replications.}
\label{tab:bch}
\end{table}

\begin{table}[p]
\centering

\begin{tabular}{|r|cc|cc|cc|cc|}
   \hline
Beta Model & \multicolumn{2}{c|}{dense} &   \multicolumn{2}{c|}{harmonic}  & \multicolumn{2}{c|}{moderately sparse}  & \multicolumn{2}{c|}{very sparse}   \\ 
  Autocovariance ($\rho$) & 0.5 & 0.9 & 0.5 & 0.9 & 0.5 & 0.9 & 0.5 & 0.9 \\
   \hline
Naive & 1.236 & 2.659 & 1.088 & 1.938 & 0.951 & 1.096 & 0.814 & 0.814 \\
  Elastic Net & 1.075 & 1.235 &  \bf 0.631 & 0.597 &  \bf 0.225 &  \bf 0.132 & 0.098 &  \bf 0.096 \\
   \hline
Approximate Balance & 1.153 &  \bf 1.125 & 1.034 & 0.994 & 0.921 & 0.717 & 0.827 & 0.569 \\
  Approx.~Residual Balance &  \bf 0.994 &  \bf 1.146 &  \bf 0.614 &  \bf 0.554 &  \bf 0.219 &  \bf 0.131 & 0.109 &  \bf 0.100 \\
   \hline
Inv.~ Propensity Weight & 1.236 & 2.645 & 1.084 & 1.932 & 0.950 & 1.091 & 0.814 & 0.813 \\
  Augmented IPW & 1.082 & 1.231 &  \bf 0.629 & 0.597 &  \bf 0.224 &  \bf 0.132 & 0.098 &  \bf 0.096 \\
   \hline
Weighted Elastic Net & 1.089 & 1.234 &  \bf 0.624 & 0.597 &  \bf 0.225 &  \bf 0.132 & 0.099 &  \bf 0.096 \\
  TMLE Elastic Net & 1.065 & 1.233 &  \bf 0.629 & 0.597 &  \bf 0.225 &  \bf 0.132 & 0.099 &  \bf 0.096 \\
   \hline
Double-Select + OLS & 1.312 & 2.659 & 1.064 & 1.938 & 0.629 & 0.204 &  \bf 0.092 &  \bf 0.097 \\
   \hline
\end{tabular}

\caption{Root-mean-squared error $\sqrt{\EE{(\htau - \tau)^2}}$ in the moderately sparse two-stage setting.
We used $n = 600$, $p = 2000$, and scaled the signal such that $\Norm{\beta}_2 = 1$.
All numbers are averaged over 400 simulation replications.}
\label{tab:ar}
\end{table}

\begin{table}[p]
\centering

\begin{tabular}{|r|ccccc|ccccc|}
   \hline
$n$ & \multicolumn{5}{c|}{400} & \multicolumn{5}{c|}{1000}  \\ 
  $p$ & 100 & 200 & 400 & 800 & 1600 & 100 & 200 & 400 & 800 & 1600 \\ 
   \hline
Naive & 1.734 & 1.738 & 1.734 & 1.736 & 1.747 & 1.724 & 1.679 & 1.706 & 1.698 & 1.720 \\
  Elastic Net & 0.446 & 0.468 & 0.492 & 0.517 & 0.540 & 0.376 & 0.380 & 0.389 & 0.401 & 0.413 \\
   \hline
Approximate Balance & 0.523 & 0.582 & 0.609 & 0.656 & 0.700 & 0.297 & 0.327 & 0.379 & 0.395 & 0.464 \\
  Approx.~Residual Balance &  \bf 0.249 &  \bf 0.276 &  \bf 0.270 &  \bf 0.295 &  \bf 0.310 &  \bf 0.168 &  \bf 0.175 &  \bf 0.176 &  \bf 0.179 &  \bf 0.194 \\
   \hline
Inv.~Propensity Weight & 1.060 & 1.081 & 1.111 & 1.154 & 1.189 & 0.831 & 0.831 & 0.874 & 0.875 & 0.940 \\
  Augmented IPW & 0.340 & 0.359 & 0.377 & 0.406 & 0.425 & 0.249 & 0.254 & 0.261 & 0.266 & 0.285 \\
   \hline
Weighted Elastic Net & 0.313 & 0.338 & 0.355 & 0.385 & 0.412 & 0.204 & 0.209 & 0.220 & 0.221 & 0.249 \\
  TMLE Elastic Net & 0.347 & 0.365 & 0.381 & 0.407 & 0.428 & 0.273 & 0.275 & 0.282 & 0.286 & 0.301 \\
   \hline
Double-Select + OLS & 0.285 & 0.292 & 0.301 & 0.320 & 0.339 & 0.250 & 0.250 & 0.246 & 0.244 & 0.246 \\
   \hline
\end{tabular}

\caption{Root-mean-squared error $\sqrt{\EE{(\htau - \tau)^2}}$ in the misspecified setting.
All numbers are averaged over 400 simulation replications.}
\label{tab:misspec}
\end{table}

\begin{table}[p]
\centering

\begin{tabular}{|rr|cc|cc|cc|}
   \hline
 &  & \multicolumn{2}{c|}{$\beta_j \propto 1\p{\cb{j \leq 10}}$}   & \multicolumn{2}{c|}{$\beta_j \propto 1/j^2$}  & \multicolumn{2}{c|}{$\beta_j \propto 1/j$} \\ 
  $n$ & $p$ & $\eta = 0.25$ & $\eta = 0.1$ & $\eta = 0.25$ & $\eta = 0.1$ & $\eta = 0.25$ & $\eta = 0.1$ \\ 
   \hline
400 & 800 & 0.95 & 0.87 & 0.97 & 0.88 & 0.87 & 0.70 \\
  400 & 1600 & 0.92 & 0.87 & 0.94 & 0.89 & 0.88 & 0.72 \\
  400 & 3200 & 0.90 & 0.82 & 0.94 & 0.86 & 0.86 & 0.71 \\
   \hline
800 & 800 & 0.94 & 0.92 & 0.97 & 0.92 & 0.95 & 0.85 \\
  800 & 1600 & 0.95 & 0.92 & 0.97 & 0.92 & 0.91 & 0.83 \\
  800 & 3200 & 0.94 & 0.90 & 0.95 & 0.91 & 0.91 & 0.79 \\
   \hline
1600 & 800 & 0.96 & 0.94 & 0.96 & 0.94 & 0.97 & 0.93 \\
  1600 & 1600 & 0.95 & 0.93 & 0.98 & 0.94 & 0.97 & 0.91 \\
  1600 & 3200 & 0.95 & 0.92 & 0.96 & 0.95 & 0.94 & 0.90 \\
   \hline
\end{tabular}

\caption{Coverage for approximate residual balancing confidence intervals as constructed in Corollary \ref{coro:inference},
with data generated as in the many cluster setting; we scaled the signal such that $\Norm{\beta}_2 = 3$.
The target coverage is 0.95.
All numbers are averaged over 1000 simulation replications.}
\label{tab:coverage}
\end{table}

\clearpage

\begin{appendix}

\section{Proofs}
\label{sec:proofs}

\subsection*{Proof of Proposition \ref{prop:main}}
First, we can write
\begin{align*}
\hmucc
&= \hatx^\top \hbeta +  \gamma^\top \p{\bycc -\bxcc\hbeta} \\
&= \hatx^\top \hbeta + \gamma^\top \bxcc \p{\beta - \hbeta} + \gamma^\top \varepsilon.
\end{align*}
Thus,
\begin{align*}
\hmucc - \mucc
&= \hatx^\top \p{\hbeta - \beta} + \gamma^\top \bxcc \p{\beta - \hbeta} + \gamma^\top \varepsilon \\
&= \p{\hatx - \bxcc^\top \gamma}^\top \p{\hbeta - \beta} + \gamma^\top \varepsilon,
\end{align*}
and so the desired conclusion follows by H\"older's inequality.
\endproof

\subsection*{Proof of Lemma \ref{lemm:OLS_balance}}
For any $j = 1, \, ..., \, p$, write
\begin{align*}
\p{\bxcc^\top \gamma^*}_j
&= \frac{1}{\nc} e_j^\top \bxcc^\top \bxcc \Sigma_{\rm c}^{-1} \xi \\
&= \frac{1}{\nc} \sum_{i} Q_i^\top A_j Q_i, \ \ A_j := \Sigma_{\rm c}^{-\frac{1}{2}} \xi e_j^\top \Sigma_{\rm c}^{\frac{1}{2}},
\end{align*}
where $e_j$ is the $j$-th basis vector, and $Q_i$ denotes the $i$-th row of the $Q$ matrix
(defined in Assumption \ref{assu:indep}) as a column vector.
Here, $A_j$ is a rank-1 matrix, with Frobenius norm
\begin{align*}
\Norm{A_j}_F^2 &= \tr\p{\Sigma_{\rm c}^{\frac{1}{2}} e_j \xi^\top \Sigma_{\rm c}^{-1} \xi e_j^\top \Sigma_{\rm c}^{\frac{1}{2}}} 
= (\Sigma_{\rm c})_{jj} \, \xi^\top \Sigma_{\rm c}^{-1} \xi \leq VS.
\end{align*}
We can now apply the Hanson-Wright inequality, as presented in Theorem 1.1 of \citet{rudelson2013hanson}.
Given our assumptions on $Q_i$---namely that it have independent, standardized, and sub-Gaussian
entries---the Hanson-Wright inequality implies that $Q_i^\top A_j Q_i$ is sub-Exponential;
more specifically, there exist universal constants $C_1$ and $C_2$ such that
$$ \EE{e^{t \p{Q_i^\top A_j Q_i - \EE{Q_i^\top A_j Q_i}}}} \leq \exp\sqb{C_1 t^2 \varsigma^4 VS} \text{ for all } t \leq \frac{C_2}{\varsigma^2 \sqrt{VS}}. $$
Thus, noting that $\EE{\bxcc^\top \gamma^*} = \xi$,
we find that for any sequence $t_n > 0$ with $t_n^2/n \rightarrow 0$,
the following relation holds for large enough $n$:
$$ \EE{\exp\sqb{ \sqrt{n} \, t_n \p{\bxcc^\top \gamma^* - \xi}_j}} \leq \exp\sqb{C_1 t_n^2 \, \varsigma^4 \,  {VS}}. $$
We can turn the above moment bound into a tail bound by applying Markov's inequality.
Plugging in $t_n := \sqrt{\log(p/2\delta)} \, / \, (\varsigma^2 \sqrt{C_1VS})$ and also applying a symmetric
argument to $(-\bxcc^\top\gamma^* + \xi)_j$, we find that for large enough $n$ and any $\delta > 0$,
$$ \PP{\abs{\sqrt{n} \, \p{X^\top \gamma^* - \xi}_j} > 2  \varsigma^2 \sqrt{C_1 VS \log\p{\frac{p}{2\delta}}}} \leq \frac{\delta}{p}. $$
The desired result then follows by applying a union bound, and noting that $\Norm{\gamma^*}_\infty \leq n^{-2/3}$ with probability tending to 1 by sub-Gaussianity of $Q_{ij}$.
\endproof

\subsection*{Proof of Theorem \ref{theo:debiased_prediction}}

We start by mimicking Proposition \ref{prop:main}, and write
\begin{equation}
\label{eq:debias_main}
\begin{split}
\htheta -\theta &= \xi \cdot \p{\hbetacc - \betacc} + \sum_{\cb{i : W_i = 0}} \gamma_i\p{Y_i  - X_i \cdot \hbetacc} \\
&= \sum_{\cb{i : W_i = 0}} \gamma_i \varepsilon_i(0) +  \p{\xi - \bxcc^\top \gamma} \cdot \p{\hbetacc - \betacc} \\
&= \sum_{\cb{i : W_i = 0}} \gamma_i \varepsilon_i(0) + \oo\p{ \Norm{\xi - \bxcc^\top \gamma}_\infty \Norm{\hbetacc - \betacc}_1}
\end{split}
\end{equation}
The proof of our main result now follows by analyzing the above bound using Lemma \ref{lemm:OLS_balance}
from the main text, as well as technical results proved below in Lemmas \ref{lemm:lasso_bound} and \ref{lemm:clt}. 

We first consider the error term. On the event that \eqref{eq:gamma_bdd} is feasible---which,
by Lemma \ref{lemm:OLS_balance} will occur with probability tending to 1---we know that
$ \Norm{\xi - \bxcc^\top \gamma}_\infty = \oo(\sqrt{\log(p)/\nc})$. Meanwhile, given our assumptions,
we can obtain an $L_1$-risk bound for the lasso that scales as $\oo(k \sqrt{\log(p)/\nc})$; see
Lemma \ref{lemm:lasso_bound}.
Taken together, these results imply that
\begin{equation}
\label{eq:error_decay}
 \Norm{\xi - \bxcc^\top \gamma}_\infty \Norm{\hbetacc - \betacc}_1 = \oo\p{\frac{k \log\p{p}}{\nc}},
\end{equation}
which, by Assumption \ref{assu:spar}, decays faster than $1/\sqrt{\nc}$.

Next, to rule out superefficiency, we need a lower bound on $\Norm{\gamma}_2^2$.
By our minimum estimand size assumption we know that there exists an index $j \in \cb{1, \, ..., \, p}$
with $\abs{\xi_j} \geq \kappa$; and thus, any feasible solution to \eqref{eq:gamma_bdd} must
eventually satisfy \smash{$(\bxcc^\top \gamma)_j^2 \geq \kappa^2/2$}. By Cauchy-Schwarz, this
implies
$$ \Norm{\gamma}_2^2 \geq \kappa^2 \,\big/\,\p{2 \sum_{\cb{i : W_i = 0}} \bx_{ij}^2} = \Theta_P\p{\frac{1}{\nc}}, $$
as desired. Given this result, a standard application of Lyapunov's central limit theorem
(Lemma \ref{lemm:clt}) paired with the bound \eqref{eq:error_decay} implies that, by
Slutsky's theorem,
$$ \p{\htheta - \theta} / \Norm{\gamma}_2^2 \Rightarrow \nn\p{0, \, \sigma^2}, $$
which was the first part of our desired conclusion.

Finally, we need to characterize the scale of the main term. To do so,
consider the weights $\gamma^*$ defined in Lemma \ref{lemm:OLS_balance}.
The concentration bound from Theorem 2.1 in \citet{rudelson2013hanson}
implies that $\nc \Norm{\gamma^*}_2 / (\xi^\top \Sigma_{\rm c}^{-1} \xi) \rightarrow_p 1$,
and so Lemma \ref{lemm:OLS_balance} implies that, with probability tending to 1, the optimization
program for $\gamma$ is feasible and
\begin{equation*}
\nc \Norm{\gamma}_2^2 \, \big/ \, \p{\xi^\top \Sigma_{\rm c}^{-1} \xi } \leq 1 + o_p(1),
\end{equation*}
thus concluding the proof.
\endproof

\begin{lemm}
\label{lemm:lasso_bound}
Under the conditions of Theorem \ref{theo:debiased_prediction}, the lasso
satisfies
\begin{equation}
\label{eq:beta_bound}
\Norm{\hbetacc - \betacc}_1 \leq \frac{5 \varsigma^2}{4} \, \frac{24 \, \upsilon}{\omega} \, k \, \sqrt{\frac{\log p}{\nc }}.
\end{equation}

\proof
Given our well-conditioning
assumptions on the covariance $\Sigma_{\rm c}$, Theorem 6 of \citet{rudelson2013reconstruction}
implies that the matrix $\bxcc$ will also satisfy a weaker restricted eigenvalue
property with high probability. Specifically, in our setting Assumption \ref{assu:spar}
implies that $\log(p) \ll \sqrt{\nc}$, and so we can use the work of \citet{rudelson2013reconstruction}
to conclude that \smash{$\nc^{-1/2} \, \bxcc$} satisfies the $\cb{k, \, \omega, \, 4}$-restricted eigenvalue condition
with high probability.

Next, given Assumption \ref{assu:indep}, we can use Theorem 2.1 of \citet{rudelson2013hanson}
to verify that the design matrix is $\bxcc$ column standardized with high probability in the sense that,
 with probability tending to 1,
$$\nc ^{-1} \sum_{\cb{i: W_i = 0}} (\bxcc)_{ij}^2 \leq (5/4)^2 \varsigma^4 S \text{ for all } j = 1, \, ..., \, p. $$
Thus, pairing these two fact about $\bxcc$ with sparsity as in Assumption \ref{assu:spar}
and the sub-Gaussianity of the noise  \smash{$\varepsilon_i(0)$}, we can use
the results of \citet{negahban2012unified} to bound the $L_1$-risk of the lasso.
Specifically, their Corollary 2 implies that, if we obtain \smash{$\hbetacc$} by running the
lasso with \smash{$\lambda = 5 \, \varsigma^2 \, \upsilon \, S\, \sqrt{\log(p)/\nc }$}, 
then, with probability tending to 1, \eqref{eq:beta_bound} holds.
Formally, to get this result, we first scale down the design by a factor $5 \varsigma^2 / 4$,
and then apply the cited result verbatim;
note that we also need to re-scale the restricted eigenvalue parameter $\omega$.
\endproof
\end{lemm}

\begin{lemm}
\label{lemm:clt}
Under the setting of Theorem \ref{theo:debiased_prediction}, suppose that
$\max_i \abs{\gamma_i} \leq \nc^{-2/3}$ and $\Norm{\gamma}_2^2 = \Omega_p(1/\nc)$.
Then, we obtain a central limit theorem
\begin{equation}
\label{eq:clt}
\frac{1}{\Norm{\gamma}_2} \sum_{\cb{i : W_i = 0}} \gamma_i \varepsilon_i(0) \Rightarrow \nn\p{0, \, \sigma^2}.
\end{equation}
\proof
The proof follows Lyapunov's method.
Since the optimization program for $\gamma$ did not consider
the outcomes $Y_i$, unconfoundedness (Assumption \ref{assu:unconf}) implies that
$\varepsilon_i(0)$ is independent of $\gamma_i$ conditionally on $X_i$, and so
$$ \EE{\sum_{\cb{i : W_i = 0}} \gamma_i \, \varepsilon_i(0) \cond \gamma} = 0 \eqand \ \Var{\sum_{\cb{i : W_i = 0}} \gamma_i \, \varepsilon_i(0) \cond \gamma} = \sigma^2 \Norm{\gamma}_2^2. $$
Next, we can again use unconfoundedness to verify that
\begin{align*}
\EE{\sum_{\cb{i : W_i = 0}} \p{ \gamma_i \, \varepsilon_i(0) }^3 \cond \gamma}
= \sum_{\cb{i : W_i = 0}} \gamma_i^3 \ \EE{\p{\varepsilon_i(0)}^3 \cond X_i} 
\leq C_3 \, \upsilon^3  \sum_{\cb{i : W_i = 0}} \gamma_i^3 \leq C_3 \, \upsilon^3 \, \nc^{-2/3} \, \Norm{\gamma}_2^2
\end{align*}
for some universal constant $C_3$, where the last inequality follows by sub-Gaussianity of $\varepsilon$ and by noting the upper bound on $\gamma_i$ in \eqref{eq:gamma_bdd}.
Thus,
$$ \EE{\sum_{\cb{i : W_i = 0}} \p{ \gamma_i \, \varepsilon_i(0) }^3 \cond \gamma} \ \bigg/ \ \Var{\sum_{\cb{i : W_i = 0}} \gamma_i \, \varepsilon_i(0) \cond \gamma}^{{3}/{2}}  = \oo\p{\nc^{-2/3} \Norm{\gamma}_2^{-1}} = o_P(1), $$
because, by assumption, \smash{$\Norm{\gamma}_2^{-1} = \oo_P\p{\sqrt{\nc}}$}.
Thus Lyapunov's theorem implies the central limit statement \eqref{eq:clt}.
\endproof
\end{lemm}

\subsection*{Proof of Corollary \ref{coro:debiased_catt}}

The key idea in establishing this result is that we need to replace the ``oracle'' weights defined
in Lemma \ref{lemm:OLS_balance} with
\begin{equation}
\label{eq:mt_weights}
\gamma^{**}_i = \frac{1}{\nc} m_{\rm t} \Sigma_{\rm c}^{-1} X_i.
\end{equation}
Once we have verified that, with high probability, these candidate weights $\gamma^{**}$ satisfy
the constraint from \eqref{eq:gamma_bdd},
i.e., $\Norm{\hatx - \bxcc^\top \gamma^{**}}_\infty \leq K \sqrt{\log(p)/\nc}$,
we can establish the result \eqref{eq:debiased_prediction_catt} by replicating the proof of
Theorem \ref{theo:debiased_prediction} verbatim.
Now, by Lemma \ref{lemm:OLS_balance}, we know that with probability tending to 1,
$$ \Norm{m_{\rm t} - \bxcc^\top \gamma^{**}}_\infty \leq C \varsigma^2  \sqrt{VS\log(p) \, \big/ \, \nc}. $$
Meanwhile, a standard Hoeffding bound together with the fact that $\nt/\nc \rightarrow_p \rho$
establishes that, with probability tending to 1,
$$ \Norm{\hatx - m_{\rm t}}_\infty \leq \nu \sqrt{2.1 \rho} \sqrt{\log(p) \, \big/ \, \nc}. $$
Combining these two bounds yields the desired result.
\endproof

\subsection*{Proof of Theorem \ref{theo:overlap}}

For concreteness, we note that we are studying the following estimator,
\begin{align}
\label{eq:gamma_bdd_overlap}
&\gamma = \argmin_{\tgamma} \cb{\Norm{\tgamma}_2^2 \ : \   \Norm{\hatx - \bxcc^\top \tgamma}_\infty \leq K \sqrt{\frac{\log(p)}{\nc}}, \, \sum_{\cb{i : W_i = 0}} \tgamma_i = 1, \, 0 \leq \tgamma_i \leq \nc^{-2/3}}, \\
\label{eq:hmu_overlap}
&\hmucc = \hatx \cdot \hbetacc + \sum_{\cb{i : W_i = 0}} \gamma_i \p{Y_i^\obs - X_i \cdot \hbetacc},
\end{align}
\smash{$K = \nu\sqrt{2.1(\rho + (\eta^{-1}-1)^2}$}.
Note that, here, we have re-incorporated the positivity and sum constraints on $\gamma$. The
positivity constraint stops us from extrapolating outside of the support of the data, and appears to improve
robustness to model misspecification.

Our proof mirrors the one used for Theorem \ref{theo:debiased_prediction}. We again start by proposing
a class of candidate weights $\gamma^*$ that satisfy the constraints \eqref{eq:gamma_bdd_overlap}; except,
this time, we motivate our candidate weights using the overlap assumption:
\begin{equation}
\label{eq:gammastar_overlap}
\gamma^*_i =  \frac{e\p{X_i}}{1 - e\p{X_i}} \, \Big/ \, \sum_{\cb{i : W_i = 0}} \frac{e\p{X_i}}{1 - e\p{X_i}}. 
\end{equation}
We start by characterizing the behavior of these weights below; we return to verify these bounds at the
end of the proof.
We also note that these weights also trivially satisfy \smash{$\gamma^*_i \leq \nc^{-2/3}$} once $\nc$ is large enough.

\begin{lemm}
\label{lemm:overlap}
Under the conditions of Theorem \ref{theo:overlap}, the weights $\gamma^*$ defined in
\eqref{eq:gammastar_overlap} satisfy the following bounds with probability at least $1 - \delta$,
for any $\delta > 0$:
\begin{align}
\label{eq:overlap_balance1}
&\Norm{\hatx - \bxcc^\top \gamma^*}_\infty \leq \nu \, \sqrt{2  \, \log\p{\frac{10 \, p}{\delta}}  \p{ \frac{1}{\nt} + \frac{\p{1 - \eta}^2}{\nc \, \eta^2} } } + \oo\p{\frac{1}{\nc}}, \ \eqand \\
\label{eq:overlap_balance2}
&\nt \, \Norm{\gamma^*}_2^2 \leq \frac{1}{\rho_n^2} \, \EE{\left.\p{\frac{e(X_i)}{1 - e(X_i)}}^2 \right| W_i = 0}
\\ \notag &\ \ \ \ \ \ \ \ \ \ \ \ \ \ \ \
+ \frac{\p{1 - \eta}^2\p{2 - 2\eta + \rho_n\eta}}{\rho_n^3\,\eta^3} \, \sqrt{\frac{1}{2\nc}  \, \log\p{\frac{10 \, p}{\delta}} } + \oo\p{\frac{1}{\nc}},
\end{align}
where  $\rho_n = \PP{W_i = 1}/\PP{W_i = 0}$ is the odds ratio for the $n$-th problem.
\end{lemm}

Given these preliminaries, we can follow the proof of Theorem \ref{theo:debiased_prediction} closely.
First, by the same argument as used to prove Lemma \ref{lemm:lasso_bound}, we can verify that
if we obtain \smash{$\hbetacc$} by running the lasso with \smash{$\lambda = 5 \, \nu \, \upsilon \, \sqrt{\log(p)/\nc }$}, 
then, with probability tending to 1,
\begin{equation}
\label{eq:beta_bound_overlap}
\Norm{\hbetacc - \betacc}_1 \leq \frac{5 \nu}{4} \, \frac{24 \, \upsilon}{\omega} \, k \, \sqrt{\frac{\log p}{\nc }}.
\end{equation}
Thus, we again find that
\begin{equation}
\label{eq:error_decay2}
\sqrt{n} \, \Norm{\hatx -\bxcc^\top \gamma}_\infty \Norm{\hbetacc - \betacc}_1 = \oo_P\p{\frac{k \, \log(p)}{\sqrt{n}}}, 
\end{equation}
and so, thanks to our sparsity assumption,
we can us Proposition \ref{prop:main} to show that the error in $\hbetacc$ does not affect the
asymptotic distribution of our estimator at the $\sqrt{n}$-scale; provided the
problem \eqref{eq:gamma_bdd_overlap}  is feasible.

Next, thanks to Lemma \ref{lemm:overlap}, we know that the problem \eqref{eq:gamma_bdd_overlap} 
is feasible with high probability. Moreover,
because the weights $\gamma$ obtained via \eqref{eq:gamma_bdd_overlap} satisfy
$\sum \gamma = 1$, we trivially find that $\Norm{\gamma}_2^2 \geq 1/\nc$ and can
apply Lemma \ref{lemm:clt} to get a central limit result for $\hmucc - \hmucc$.
Finally, invoking \eqref{eq:overlap_balance2} and the fact that $\Norm{\gamma}_2 \leq \Norm{\gamma^*}_2$
with probability tending to 1, we obtain the desired rate bound \eqref{eq:inference_rate}.
\endproof

\subsubsection*{Proof of Lemma \ref{lemm:overlap}}
 
To verify our desired result, first note that because $\sum \gamma^*_i = 1$, our main quantity of interest $\hatx - \bxcc \gamma^*$ is translation invariant (i.e., we can map $X_i \rightarrow X_i + c$ for any $c \in \RR^p$ without altering the quantity). Thus, we can without loss of generality re-center our problem such that $\EE{X_i \cond W_i = 1} = 0$. Given this re-centering, we use standard manipulations of sub-Gaussian random variables to check that, conditionally on $\nc $ and $\nt $ and for every $j = 1, \, ... ,\, p$:
\begin{itemize}
\item $\hatxj = \nt ^{-1} \sum_{\cb{i : W_i = 1}} X_{ij}$ is sub-Gaussian with parameter $\nu^2/\nt $ by sub-Gaussianity of $X_{ij}$ as in Assumption \ref{assu:tech}.
\item $A_j := \nc ^{-1} \sum_{\cb{i : W_i = 0}} X_{ij} \, e(X_i)/ (1 - e(X_i))$ is sub-Gaussian with parameter $\nu^2 (1 - \eta)^2 / (\nc  \, \eta^2)$ by sub-Gaussianity of $X_{ij}$ and because $e(X_i) \leq 1 - \eta$. Note that, by construction $\EE{A_j} = \EE{X_j \cond W = 1}$, and so given our re-centering $\EE{A_j} = 0$.
\item $D := \nc ^{-1} \sum_{\cb{i : W_i = 0}} e(X_i)/ (1 - e(X_i)) - \rho_n$ is sub-Gaussian with parameter $(1 - \eta)^2 / (4\nc  \,  \eta^2)$, where $\rho_n = \PP{W = 1}/\PP{W=0}$ denotes the odds ratio.
\item $V := \nc ^{-1} \sum_{\cb{i : W_i = 0}} (e(X_i)/ (1 - e(X_i))^2$ is sub-Gaussian with parameter $(1 - \eta)^4 / (4\nc  \,  \eta^4)$ after re-centering.
\end{itemize}
 Next, we apply a union bound, by which, for any $\delta > 0$, the following event $\ee_\delta$ occurs with probability at least $1 - \delta$:
 \begin{align*}
 & \Norm{A}_\infty \leq \nu\, \p{1 - \eta} \,/ ( \eta \, \sqrt{\nc }) \, \sqrt{2 \log(10 \, p \, \delta^{-1})}, \  \\
 & \Norm{\hatx - A}_\infty \leq \nu\, \sqrt{1/\nt  + \p{1 - \eta}^2 \,/\, (\nc  \, \eta^2)} \, \sqrt{2 \log(10 \, p \, \delta^{-1})}, \  \\
 &\abs{D} \leq \p{1 - \eta} \,/ (2 \eta \, \sqrt{\nc }) \, \sqrt{2 \log(10 \, \delta^{-1})}, \eqand \\
 & V \leq \EE{V} +  \p{1 - \eta}^2 \,/ (2 \eta^2 \, \sqrt{\nc }) \, \sqrt{2 \log(10 \, \delta^{-1})}.
 \end{align*}
 We then see that on the event $\ee_\delta$,
 \begin{align*}
\Norm{\hatx - \bxcc^\top \gamma^*}_\infty &= \Norm{\hatx - \p{\rho_n + D}^{-1} A}_\infty \leq \Norm{\hatx - A}_\infty + \abs{\frac{D}{\rho_n + D}} \, \Norm{A}_\infty \\
&\leq \nu\, \sqrt{\frac{1}{\nt } + \frac{\p{1 - \eta}^2}{\nc  \, \eta^2}} \ \sqrt{2 \log\p{\frac{10 \, p}{\delta}}} + \oo\p{\frac{1}{\nc }}.
 \end{align*}
 Moreover, noting that
 $$\EE{V} = \EE{\frac{e(X_i)^2}{\p{1 - e(X_i)}^2} \cond W_i = 0} \leq \frac{(1 - \eta)^2}{\eta^2}, $$
 we see that on $\varepsilon_\delta$,
 \begin{align*}
 \nc  \Norm{\gamma^*}_2^2 = \frac{V}{\p{\rho_n + D}^2} \leq \frac{\EE{V}}{\rho_n^2} + \p{\frac{1}{2}  +  \frac{1 - \eta}{\rho_n \, \eta}} \ \frac{\p{1 - \eta}^2}{\rho_n^2 \eta^2} \ \sqrt{\frac{2}{\nc } \log\p{\frac{10}{\delta}}} + \oo\p{\frac{1}{\nc }},
 \end{align*}
and so $\gamma^*$ in fact satisfies all desired constraints.
\endproof

\subsection*{Proof of Corollary \ref{coro:inference}}

We prove the result in the setting of Theorem \ref{theo:overlap}.
First of all, we can use the argument of Theorem \ref{theo:overlap} verbatim to show that
$$\p{\hmucc - \mucc} \,\Big/\, \sqrt{V_{\rm c}} \Rightarrow \nn\p{0, \, 1}, \ \ V_{\rm c} = \sum_{\cb{i : W_i = 0}} \gamma_i^2 \ \Var{\varepsilon_i(0) \cond X_i}. $$
To establish this claim, note that our bias bound \eqref{eq:error_decay2}
did not rely on homskedasticity, and the Lyapunov central limit theorem remains valid as long
as the conditional variance of \smash{$\varepsilon_i(0)$} remains bounded from below.
Thus, in order to derive the pivot \eqref{eq:pivot}, we only need to show that \smash{$\hV_{\rm c}/V_{\rm c} \rightarrow_p 1$};
the desired conclusion then follows from Slutsky's theorem.
Now, to verify this latter result, it suffices to check that
\begin{align}
\label{eq:limit1}
&\frac{1}{V_{\rm c}} \sum_{\cb{i : W_i = 0}} \gamma_i^2 \p{Y_i - X_i \cdot \betacc}^2 \rightarrow_p 1, \eqand \\
\label{eq:limit2}
&\frac{1}{V_{\rm c}} \sum_{\cb{i : W_i = 0}} \gamma_i^2  \p{X_i \cdot \p{\betacc - \hbetacc}}^2 \rightarrow_p 0.
\end{align}
To show the first convergence result, we can proceed as in the proof of Lemma \ref{lemm:clt} to verify that there is a universal constant $C_4$ for which
\begin{align*}
\Var{\sum_{\cb{i : W_i = 0}} \gamma_i^2 \p{Y_i - X_i \cdot \betacc}^2 \cond \gamma}
\leq  C_4  \, \upsilon^4 \, \Norm{\gamma}_4^4
\leq  C_4  \, \upsilon^4 \, \nc^{-4/3} \, \Norm{\gamma}_2^2,
\end{align*}
and so \eqref{eq:limit1} holds by Markov's inequality. 
\sloppy{Meanwhile, to establish \eqref{eq:limit2}, we focus on the case $\liminf \log(p) / \log(n) > 0$.}
We omit the argument in the ultra-low dimensional case since, when $p \ll n^{0.01}$, there is no strong reason to run our method instead of classical methods based on ordinary least squares.
Now, we first note the upper bound
\begin{align*}
\sum_{\cb{i : W_i = 0}} \gamma_i^2  \p{X_i \cdot \p{\betacc - \hbetacc}}^2
&\leq \Norm{\gamma}_2^2 \, \Norm{\bxcc \p{\betacc - \hbetacc}}_\infty^2
\leq \Norm{\gamma}_2^2 \, \Norm{\bxcc}_\infty^2 \Norm{\betacc - \hbetacc}_1^2,
\end{align*}
where the second step uses H\"older's inequality as in the proof of Proposition \ref{prop:main}.
Then, thanks to the assumed upper and lower bounds on the conditional variance of $\varepsilon_i(W_i)$
given $X_i$ and $W_i$, we only need to check that
$$ \Norm{\bxcc}_\infty^2 \Norm{\betacc - \hbetacc}_1^2 \rightarrow_p 0. $$
We can use sub-Gaussianity of $X_i$ (Assumption \ref{assu:tech}) and the bound
\eqref{eq:beta_bound_overlap} on the $L_1$-error of $\hbetacc$ to find a constant
$C(\nu, \, \omega, \, \upsilon)$ for which
$$  \Norm{\bxcc}_\infty^2 \Norm{\betacc - \hbetacc}_1^2 \leq C(\nu, \, \omega, \, \upsilon) \, \log\p{p \, \nc} \ k^2 \, \frac{\log\p{p}}{\nc} $$
with probability tending to 1. Then, noting our sparsity condition on $k$ (Assumption \ref{assu:spar}), we find that
$$ \log\p{p \, \nc} \ k^2 \, \frac{\log\p{p}}{\nc} \ll \frac{\log\p{p \, \nc}}{\log(p)}, $$
which is bounded from above whenever $\liminf \log(p) / \log(n) > 0$.

\end{appendix}

\end{document}